\documentclass[prb]{revtex4-2}
\usepackage{graphicx}
\usepackage{dcolumn}
\usepackage{bm}
\usepackage [latin1]{inputenc}
\usepackage{epstopdf}
\usepackage{color}
\usepackage{ulem}
\usepackage{multirow}
\begin{document}

\title{{Control of charge state of dopants
in insulating crystals: case study of Ti-doped sapphire.}}

\author{L.\,Yu.\,Kravchenko$^1$, D.\,V.\,Fil$^{1,2}$}

\email{fil@isc.kharkov.ua}

\affiliation{$^1$Institute for Single Crystals, National Academy of Sciences of Ukraine,
60 Nauky Avenue, Kharkiv 61072, Ukraine\\
$^2$V.N. Karazin Kharkiv National University, 4 Svobody Square, Kharkiv 61022, Ukraine}

\begin{abstract}

We study mechanisms of control of charge state and concentration of different point
defects in  doped insulating crystals. The approach is based on the density functional
theory calculations. We apply it to the problem of obtaining of Ti-doped sapphire
crystals with high figure-of-merit (FOM). The FOM of a given  sample  is defined as the
ratio of the coefficient of absorption  at the pump frequency to the coefficient of
absorption at the working frequency of Ti:sapphire laser. It is one of standard
specifications of commercial Ti:sapphire  laser crystals. It is believed that FOM is
proportional to the ratio of the concentration of isolated Ti$^{3+}$ ions to the
concentration of Ti$^{3+}$-Ti$^{4+}$ pairs. We find that generally this ratio is in
inverse proportion to the concentration of Ti$^{4+}$ isolated substitutional defects with
the coefficient of proportionality that depends on the temperature at which the
thermodynamically equilibrium concentration of defects is reached. We argue that in
certain cases the inverse proportion between concentrations of Ti$^{3+}$-Ti$^{4+}$ and
Ti$^{4+}$
 may be violated. The role of codoping in the control of the charge state of
dopants is analyzed. We show that codopants that form positively (negatively) charged
defects may decrease (increase) the concentration of positively charged defects formed by
the main dopants. To evaluate the effect of codoping it is important to take into account
not only isolated defects but defect complexes formed by codopants, as well.  In
particular, we show that codoping of Ti:sapphire with nitrogen results in an essential
increase of the concentration of Ti$^{4+}$ and in a decrease of the FOM, and,
consequently, growth or annealing in the presence of nitrogen or its compounds is
unfavorable  for producing Ti:sapphire laser crystals. The approach developed can be used
for determining appropriate growth and annealing conditions for obtaining doped crystals
with the required characteristics.
   \end{abstract}

\maketitle

\section{Introduction}

Doping of insulating crystals with active ions is a widely used method of obtaining
functional materials (active laser medium, luminescent materials, scintillators and many
others) with required properties. Usually, in such materials dopant ions should be a
certain charge state, occupy certain crystallography positions, and not form (or,
instead, form) complexes with other dopant ions or intrinsic defects. The problem of
control of the state of the dopant in a host matrix remains very important.

Density functional theory (DFT) is a powerful tool for evaluation of efficiency of
different methods of such  control. Using the results of DFT calculations and
thermochemical data one can analyze finite-temperature properties. In particular, one can
calculate equilibrium concentrations of different defects at a given temperature.  Due to
the condition of overall charge neutrality of defects the concentration of  a given
charged defect species cannot be calculated independently: in the general case this
concentration depends on formation energies of all charged defects. One can imply that
equilibrium (or almost equilibrium) concentrations of defects are reached under
annealing. Therefore the temperature that enters into equations for equilibrium
concentrations of defects can be associated with the temperature of annealing. For
as-grown samples it can be replaced with the melting temperature $T_m$. The situation
becomes more complicated  in a situation where concentration of dopants that enter into
the crystal under its growth is smaller than their equilibrium concentration at $T=T_m$.
In this case the concentration of a given specie depends on the formation energies of all
charged and uncharged defects.

In this paper we consider the problem of control of the charge state of dopants with
reference to Ti-doped sapphire crystals. Some elements of our approach were already
presented in our previous paper \cite{my}, where the results of DFT study of defect
complexes of Ti-doped sapphire were reported.

We start with a short introduction where we describe the problem as it is formulated in
the material science community.

Ti-sapphire is a widely used active laser medium. Operation of a Ti:Al$_2$O$_3$ tunable
laser was first reported by Moulton \cite{m} in 1982. An active ion in Ti:sapphire is
Ti$^{3+}$ substituted for the octahedrally coordinated Al$^{3+}$. This ion has a single
3d electron above a closed shell. Five d-electron levels are split by the crystal field
into an $e_g$ doublet and $t_{2g}$ triplet. Transitions between the $t_{2g}$ and $e_g$
levels are responsible for the absorption of visible light and near-infrared fluorescence
\cite{1,2}. Titanium  substituted for the Al ion can also be in the  Ti$^{4+}$ state. A
charge-transfer transition between O$^{2-}$ and Ti$^{4+}$ causes ultraviolet (UV)
absorption in Ti:sapphire \cite{uv}. Measurement of UV spectral characteristics gives the
information on the concentration of Ti$^{4+}$
 ions  in Ti-sapphire samples \cite{uv, nizh}.

 Ti:sapphire   exhibits weak
near-infrared (NIR) absorption \cite{1,2,2a,3a,3,4} that results in losses at the
wavelength of the laser emission. To qualify the performance of Ti:sapphire as a laser
crystal the ratio of the absorption $\alpha_m$ at the pump wavelength ($\lambda_m\approx
500$ nm) to the absorption $\alpha_r$ at the laser emission wavelength ($\lambda_r\approx
800$ nm) is used.  This ratio is known as a figure-of-merit (FOM)  characterization of
commercial materials.  To calculate the FOM one should specify exact values of
$\lambda_m$ and $\lambda_r$. One of the accepted choices
 is $\lambda_m=514$ nm and $\lambda_r=820$ nm \cite{pinto}. Slightly different
 $\lambda_m$ and $\lambda_r$ are also accepted \cite{3}.

 NIR absorption is associated with  Ti$^{3+}$ -Ti$^{4+}$
pairs. A correlation between NIR absorption and the concentration of  Ti$^{3+}$-Ti$^{4+}$
 pairs
 was demonstrated in \cite{3} where the dependence of $\alpha_{r}$ on
$\alpha_{m}$ was measured. A partially oxidized sample studied in
 \cite{3} was clear near the surface and pink inside that visualized a
  variation of the concentration of Ti$^{3+}$ across the sample.
  The total concentration of Ti $c_{\mathrm{Ti}}$ was a constant.
 The obtained dependence of $\alpha_{r}$ on $\alpha_{m}$ is of a bell-like
shape. This dependence is described by the formula $\alpha_{r}\propto
\alpha_{m}(\alpha_0-\alpha_{m})$, where $\alpha_0$ is some constant. It is believed that
 the concentration of $\mathrm{Ti}^{3+}$-$\mathrm{Ti}^{4+}$ pairs
 ($c_{\mathrm{Ti}^{3+}-\mathrm{Ti}^{4+}}$) is
proportional to the product of concentrations of isolated $\mathrm{Ti}^{3+}$ and
$\mathrm{Ti}^{4+}$ ions. The first one is proportional to $\alpha_m$, and the second, to
$c_\mathrm{Ti}-c_{\mathrm{Ti}^{3+}}\propto \alpha_0-\alpha_{m}$. Therefore the
observation \cite{3} correlates with the expectation that the adsorption coefficient
$\alpha_{r}$ depends linearly on $c_{\mathrm{Ti}^{3+}-\mathrm{Ti}^{4+}}$. The pair
mechanism of NIR absorption is supported by calculations  of energy levels of the
$\mathrm{Ti}^{3+}\mathrm{Ti}^{4+}\mathrm{O}_9^{2-}$ cluster \cite{zha}.
 The energy
difference \cite{zha} between the ground state and the first excited
level of such a cluster corresponds to the wavelength $\lambda=813$
nm  that is in excellent agreement with the experimental value of
$\lambda$ at which the maximum of NIR absorption is observed.

The problem of NIR absorption was revisited recently in \cite{m1} where the absorption
data in the range of wavelengths from 190 nm to 2000 nm were analyzed with reference to
many Ti:sapphire samples of different origin. A sample with a large fraction of Ti$^{4+}$
(that was confirmed by a very weak absorption by this sample at $\lambda=490$ nm) was
used as a reference. The Ti$^{4+}$ scaling factor was defined as the absorption at
$\lambda=225$ nm by a given sample related to the absorption by the reference sample. It
was implied that this factor is proportional to the concentration of Ti$^{4+}$. If NIR is
caused by $\mathrm{Ti}^{3+}$-$\mathrm{Ti}^{4+}$ pairs, the FOM will be in inverse
proportion to the Ti$^{4+}$ scaling factor. Nevertheless, some samples demonstrated
strong deviation from this law. Based on this observation the authors of Ref. \cite{m1}
arrived at the conclusion that the $\mathrm{Ti}^{3+}$-$\mathrm{Ti}^{4+}$ pair model of
NIR absorption needs a revision.

Defect energetics in Ti:sapphire were investigated within the DFT approach in
\cite{d1,d2}. It was shown that in the oxidized conditions Ti$^{4+}$ ions  substituted
for Al$^{3+}$ ions together with charge compensating vacancies of Al
($V_{\mathrm{Al}}^{3-}$) are the most stable defects. In the reduced conditions the
formation energy of substitutional Ti$^{3+}$ ions is of the smallest value. In the
intermediate range of the oxygen potential the substitutional Ti$^{3+}$ and Ti$^{4+}$
defects exhibit similar formation energies, indicating that they can coexist. It was
established that Ti$^{3+}$ ions demonstrate a tendency to form pairs and larger clusters.
The binding energy of the $\mathrm{Ti}^{3+}$-$\mathrm{Ti}^{3+}$ pair strongly depends on
the distance between Ti$^{3+}$ ions. For the first nearest neighbors this energy is about
1.2 eV, but for the third nearest neighbors it is less than 0.2 eV. The binding energy of
Ti$^{3+}$  triples and quadruples is about 2 eV and 3 eV, correspondingly. In addition,
positively charged Ti$^{4+}$ ions bind in pairs with negatively charged Al vacancies
$V_\mathrm{Al}^{3-}$ with almost the same binding energy as one of
$\mathrm{Ti}^{3+}$-$\mathrm{Ti}^{3+}$ pairs. Energetics of defect complexes in Ti-doped
sapphire was studied in detail in Ref. \cite{my}. The formation energies and the binding
energy of  pairs, triples and quadruples formed by $\mathrm{Ti}^{3+}$, $\mathrm{Ti}^{4+}$
and $V_\mathrm{Al}^{3-}$ were calculated.  It was shown that equilibrium concentrations
of complex defects
 can be on the same order of or even
larger than the concentration of isolated $\mathrm{Ti}^{3+}$ or $\mathrm{Ti}^{4+}$
defects.  It was found that complex defects in Ti:sapphire influence significantly the
balance between charged defects.

In this paper we present a thorough investigation of the problem of control of the charge
state of titanium in Ti:sapphire. Our investigation compliments the study \cite{d1,d2}
and our recent study \cite{my}.

The method for the calculation of defect formation energies and of dopant concentration
presented in Secs. \ref{s3}-\ref{s5} is the central result of this paper. It can be
applied to dopant and defect complexes in  different technologically relevant materials.
In Secs. \ref{s5} and \ref{s6} we  answer a number of questions that arose in \cite{m1}
and clarify the role of several factors that may increase the FOM such as an appropriate
choice of the temperature of annealing,  adding of certain compounds into the atmosphere
or into the melt under the growth or annealing, co-doping with nonisovalent atoms. This
analysis is another principal result of the paper. The approach developed is quite
general and can be used for determining appropriate growth and annealing conditions to
control the  charge state of dopants in various insulating crystals not restricted to
Ti:sapphire.

\section{Equilibrium concentration of defects and universal relation between
concentrations of
simple and complex defects} \label{s2}

In this section we derive the general relation between equilibrium concentrations of
isolated and complex defects and show that this relation also takes place if the total
number of dopant atoms is fixed.

Equilibrium concentrations of defects can be obtained from the condition of the minimum
of the free energy. The free energy depends on the defect formation energies $E_i$ and
the configurational entropy $\ln W$, where $W$ is the number of ways to place defects  in
the crystal:
\begin{equation}\label{1}
    F=F_0+\sum_{i}  E_i n_i -k_B T \ln W.
\end{equation}
In Eq. (\ref{1}) $F_0$ is the free energy of the perfect crystal, the sum is taken over
all possible defect species, $n_i$ is the number of defects of the $i$-th specie,
 $T$ is the temperature, and $k_B$ is the Boltzmann constant. We imply that the total number of
defects $n_{tot}=n_{i_1}+n_{i_2}+\ldots$ is much smaller than the number of lattice sites
and approximate $W$ by the equation
\begin{equation}\label{0}
  W \approx \prod_{i}  \frac{N_i!}{(N_i-n_i)! n_i!},
\end{equation}
where $N_i$ is the number of different positions and orientations  for the $i$-th defect
specie.

We take into account that overall electric charge of all charged defects is zero. This
results in the constraint
\begin{equation}\label{3}
    \sum_{i} q_i n_i=0,
\end{equation}
where $q_i$ is the electrical charge of the $i$-th defect (below we use elementary charge
units for $q_i$).  If the total number of dopant in the crystal is fixed we have the
additional constraint
\begin{equation}\label{6}
   \sum_i k_{\mathrm{Ti},i}n_i=n_{\mathrm{Ti}}.
\end{equation}
Here we specify the case of Ti dopants. In Eq. (\ref{6}) $k_{\mathrm{Ti},i}$ is the
number of Ti atoms per an $i$-th defect and $n_{\mathrm{Ti}}$ is the total number of Ti
ions in the crystal. The constraint (\ref{6}) should be taken into account if an actual
concentration of dopant differs from its equilibrium concentration at $T=T_m$ (for
as-grown samples) or at the temperature of annealing.

The free energy minimum with the additional constraints can be found by the method of
Lagrange multipliers. The constraints (\ref{3}) and (\ref{6}) can be taken into account
by two additional terms with two Lagrange multipliers $\lambda_q$ and
$\lambda_\mathrm{Ti}$. The extremum condition yields
\begin{equation}\label{7}
   \tilde{n}_i=\frac{n_i}{N_i}=
   \exp\left(-\frac{E_i- \lambda_q q_{i}-\lambda_{\mathrm{Ti}}k_{\mathrm{Ti},i}}{k_B
   T}\right).
\end{equation}
Substituting Eq. (\ref{7}) into Eqs. (\ref{3}) and (\ref{6}) we obtain two equations for
 $\lambda_q$ and $\lambda_{\mathrm{Ti}}$.

 One can find the general relation between the
concentrations of isolated and complex defects using Eq. (\ref{7}).
Let us consider a complex defect $i_c$ composed of $r_1+r_2+\ldots
r_s$ simple defects:
\begin{equation}\label{3-1}
i_c=\underbrace{i_1\ldots i_1}_{r_1\, \mathrm{times}} \underbrace{i_2\ldots i_2}_{r_2\,
\mathrm{times}}\ldots\underbrace{i_s\ldots i_s}_{r_s\, \mathrm{times}}.
\end{equation}
For such a defect Eq. (\ref{7}) can be rewritten as
\begin{eqnarray}\label{9}
    \tilde{n}_{i_c}=\exp\left(\frac{E_{i_c}^{(b)}}{k_B T}\right)\prod_{j=1}^s\left[
    \exp\left(-\frac{{E}_{i_j}- \lambda_q q_{i_j}-
     \lambda_\mathrm{Ti} k_{\mathrm{Ti},i_j}}{k_B T}\right)
    \right]^{r_j}\cr\times\exp\frac{\lambda_q\left(q_{i_c}-\sum_{j=1}^s r_{j}
 q_{i_j}\right)}{k_B T}
  \exp\frac{\lambda_\mathrm{Ti}\left(k_{\mathrm{Ti},i_c}-\sum_{j=1}^s r_{j}
 k_{\mathrm{Ti},i_j}\right)}{k_B T},
\end{eqnarray}
where
\begin{equation}\label{5}
    E^{(b)}_{i_c}=\sum_{j=1}^s r_{j} E_{i_j}-E_{i_c}
\end{equation}
is the binding energy of the complex defect $i_c$. Charge and particle number
conservation requires $q_{i_c}=\sum_{j=1}^s r_{j} q_{i_j}$
 and $k_{\mathrm{Ti},i_c}=\sum_{j=1}^s r_{j} k_{\mathrm{Ti},i_j}$. Thus  Eq. (\ref{9})
 reduces to
 \begin{equation}\label{4}
 \tilde{n}_{i_c}=\exp\left(\frac{E_{i_c}^{(b)}}{k_B T}\right)
\left(\tilde{n}_{i_1}\right)^{r_1}\left(\tilde{n}_{i_2}\right)^{r_2}\ldots
\left(\tilde{n}_{i_s}\right)^{r_s}.
\end{equation}
 Considering the problem with only one constraint (\ref{3}) we put
$\lambda_\mathrm{Ti}=0$ from the beginning and again arrive at the
relation (\ref{4}).

In the DFT approach the defect formation energy is given by the equation \cite{z-n,frey}
\begin{equation}\label{10}
    E_i=E_{\mathrm{def},i}-E_{\mathrm{perf}}-\sum_X \mu_X p_{X,i}+\mu_e  q_i+E^{(c)}_i,
\end{equation}
where $E_{\mathrm{def},i}$ is the energy of the supercell with a given defect,
$E_{\mathrm{perf}}$ is the energy of the perfect supercell,  $p_{X,i}$ is the number of
atoms of type X (host or impurity atoms) that have been added to ($p_{X,i}> 0$) or
removed from ($p_{X,i}<0$) the supercell to form the defect of the $i$-th species,
 $\mu_X$ is the chemical potential of the atom of the type X, $\mu_e$ is the
 electron chemical potential, and $E_i^{(c)}$ is the
correction that excludes electrostatic interaction caused by periodic coping of charged
defects in the supercell calculations.

 Substituting Eq. (\ref{10}) into Eq. (\ref{7}) and redefining the
Lagrange multiplier  $\tilde{\lambda}_q=\lambda_q-\mu_e$ one can exclude the electron
chemical potential from the problem. This means that equilibrium concentrations of
defects can be expressed through the quantities independently of $\mu_e$. If the total
number of Ti atoms is fixed, the chemical potential of Ti can be excluded as well. In the
latter case equilibrium concentrations of defects do not depend on $\mu_\mathrm{Ti}$.

Applying Eq. (\ref{4})  to Ti:sapphire  we find that the ratio of the concentration of
isolated $\mathrm{Ti}^{3+}$ ions to the concentration of
$\mathrm{Ti}^{3+}-\mathrm{Ti}^{4+}$ pairs is in inverse proportion to the concentration
of isolated   $\mathrm{Ti}^{4+}$ ions:
\begin{equation}\label{13}
  \frac{c_3}{c_{3-4}}=
  \frac{ e^{-\frac{E^{(b)}_{3-4}}{k_B T}}}{2 \tilde{n}_4} .
   \end{equation}
Here and below we use the notations  $3\equiv \mathrm{Ti}^{3+}\equiv
\mathrm{Ti}_\mathrm{Al}^0$, $4\equiv \mathrm{Ti}^{4+}\equiv
\mathrm{Ti}_\mathrm{Al}^+$, and
$3-4\equiv\mathrm{Ti}^{3+}$-$\mathrm{Ti}^{4+}$.
 The factor of 2 in the denominator of Eq. (\ref{13}) is the number of
different positions and orientations of $\mathrm{Ti}^{3+}$-$\mathrm{Ti}^{4+}$ pairs per
one Al site. The coefficient of proportionality depends on the temperature at which the
equilibrium concentration of defects is reached. For  defects with positive binding
energy (DFT calculations show that $E^{(b)}_{3-4}>0$) the coefficient of proportionality
decreases under lowering in this temperature. The relation (\ref{13}) is fulfilled for
any total concentration of Ti (equilibrium at given temperature or fixed to a certain
level). At the same time the relation (\ref{13}) may be violates under additional
constraints on the concentrations of defects (see, Sec. \ref{s6}).

\section{Calculation of defect formation energies}
\label{s3}

To calculate defect formation energies we use the Kohn-Sham density functional method in
the generalized gradient approximation with the Perdew-Burke-Ernzerhof parametrization
for the exchange-correlation functional and double-zeta basis with polarization orbitals
as implemented in the open source SIESTA code \cite{si}. The pseudopotentials were
generated with the improved Troullier-Martins scheme. The $\mathrm{Al}$-$3s^23p^1$,
$\mathrm{O}$-$ 2s^22p^4$ and $\mathrm{Ti}$-$ 4s^13d^3$ electronic states are considered
as valence ones. We do not include 3d semicore states into the valence ones due to the
following reason. The heats of formation of crystal phases used as references  for
building the Al-Ti-O phase diagram (TiO$_2$, Ti$_3$O$_5$, Ti$_2$O$_3$) calculated with
the use of the 4-electron state of Ti are in better coincidence with experimental
quantities than the ones of the 10-electron state of Ti. In addition, calculations with
10 valence electrons per atom are slower than with 4. Actually there is no unambiguous
answer as to where semicore
 states should be included \cite{pp}.
To study  co-doping with carbon, nitrogen and fluorine we set $\mathrm{C}$-$ 2s^22p^2$,
$\mathrm{N}$-$ 2s^22p^3$, and $\mathrm{F}$-$ 2s^22p^5$ valence electronic states.

Lattice parameters and atomic positions are optimized until the residual stress
components converge to less than 0.1GPa and the residual forces  are less than 0.01
eV/$\AA$. The plane-wave cutoff energy of 250 Ry is used to calculate the total energy of
the system. A 120-atom supercell of $\alpha $-Al$_2$O$_3$ is built of 4 optimized unit
cells. One simple or complex defect is placed in the supercell. The optimization of
atomic positions in the supercell with a defect is fulfilled again. Numerical
integrations over the supercell Brillouin zone are performed at the $\Gamma$ point. To
check convergency of the $\Gamma$ point integration we calculate several formation
energies integrating over $3\times 3\times 3$ and $5\times 5 \times 5 $ grids of
$k$-points. Integration over $3\times 3\times 3$ grid yields  Ti$^{3+}$, Ti$^{4+}$ and
$V_{\mathrm{Al}}^{3-}$ formation energies higher than ones for the $\Gamma$ point
integration by  0.02 eV, 0.027 eV and 0.018 eV, correspondingly. The difference of
energies given by integration over $3\times 3\times 3$ and $5\times 5 \times 5 $ grids of
$k$-points is less than $ 10^{-3}$ eV. Based on this results we conclude that restriction
with integration over the $\Gamma$ point (which considerably speed up the calculations)
mainly results in  unessential overestimate (about 10 \%) of the total equilibrium
concentration of Ti. The error in relative concentrations of different defects is less
than 3 percents.

Chemical potentials of atoms that enter into Eq. (\ref{10}) are calculated from chemical
potentials of materials related to the $\mathrm{Al}-\mathrm{Ti}-\mathrm{O}$ system. The
chemical potential of a given material is the sum of the zero-temperature DFT energy and
a temperature correction part:
\begin{equation}\label{18-0}
 \mu(X)=\mu_{0}(X)+\Delta \mu_T(X),
\end{equation}
where $\mu_0(X)$ is the DFT energy  of a crystal (per formula unit), or of an isolated
molecule if the material is a gas at the standard conditions. The temperature correction
is determined by the equation \cite{t-r1,t-r2}
\begin{equation}\label{18}
    \Delta \mu_T(X)=H_{X}(T,p)-H_{X}(0,p) -T S_{X}(T,p),
\end{equation}
where  $H_{X}(T,p)$ and  $S_{X}(T,p)$ are the enthalpy and entropy at the temperature $T$
and pressure $p$. We put $p=0.1$ MPa (standard conditions) and take the values of
$H_{X}(T,p)$ and $S_{X}(T,p)$ from the thermochemical tables \cite{tct}.

\begin{figure}[h]
     \center{\includegraphics[width=0.4\textwidth]{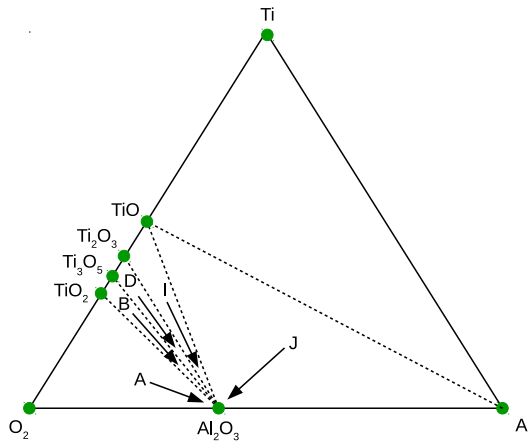}}
     \caption{Phase diagram of the Al$-$Ti$-$O system at $T=T_m$. The points A, B, D, I and
     J    denoted by arrows correspond to the vertices of
the three-phase region around Al$_2$O$_3$.}\label{f1}
     \end{figure}

The heat of formation $H_{f}$ for a compound  with the general formula ${X}_u{Y}_v$,
where ${X},{Y}=\mathrm{Al},\mathrm{Ti},\mathrm{C}$ is calculated as
\begin{equation}\label{19-1}
    H_{f}({X}_u{Y}_v)=\frac{\mu({X}_u{Y}_v
    )
    -u \mu({X})-v\mu({
Y})}{u+v}
\end{equation}
or, if ${Y} =\mathrm{O},\mathrm{N},\mathrm{F}$, as
\begin{equation}\label{19-2}
    H_{f}({X}_u{Y}_v)=\frac{\mu({X}_u{Y}_v
    )
    -u \mu({X})-\frac{v}{2}\mu({
Y}_2)}{u+v}.
\end{equation}
Equations (\ref{19-1}) and (\ref{19-2}) give heats of formation per one atom. Chemical
potentials in Eqs. (\ref{19-1}) and (\ref{19-2}) take into account the temperature
correction Eq. (\ref{18}). Using calculated heats of formation (\ref{19-1}) and
(\ref{19-2}) we build the $\mathrm{Al}-\mathrm{O}-\mathrm{Ti}$ phase diagram and define
points in which $\alpha - \mathrm{Al}_2\mathrm{O}_3$ is in equilibrium with two other
phases. The obtained phase diagram at $T=T_m=2327$K (the melting temperature of
$\alpha$-Al$_2$O$_3$) is shown  in Fig. \ref{f1}. The same procedure is done for
four-component systems $\mathrm{Al}-\mathrm{O}-\mathrm{Ti}-X$, where
$X=\mathrm{C},\mathrm{N},\mathrm{F}$. The reference points in the phase diagrams, where
two (three) phases are in equilibrium with $\mathrm{Al}_2\mathrm{O}_3$  are listed in
Table \ref{t1}.

\begin{table}
      \caption{Phases in equilibrium with
       $\mathrm{Al}_2\mathrm{O}_3$ and the oxygen chemical potential
       in reference points of the ternary and four-component phase
       diagrams       at $T=T_m$.
      The  potential $\mu_\mathrm{O}$ is counted from the energy
      of an isolated oxygen atom.}
      \label{t1}
      \begin{center}
\begin{tabular}{|c|c|c|c|c|c|}
  \hline
  Point & $\mathrm{Al}-\mathrm{O}-\mathrm{Ti}$ &
  $\mathrm{Al}-\mathrm{O}-\mathrm{Ti}-\mathrm{C}$ & $\mathrm{Al}-\mathrm{O}-\mathrm{Ti}-\mathrm{N}$
  & $\mathrm{Al}-\mathrm{O}-\mathrm{Ti}-\mathrm{F}$  & $\mu_\mathrm{O}, eV$ \\ \hline
  A &O$_2$, TiO$_2$ & O$_2$, TiO$_2$, CO$_2$ & O$_2$, TiO$_2$, N$_2$ & O$_2$, TiO$_2$, AlF$_3$ & -7.29 \\
  B & TiO$_2$,  Ti$_3$O$_5$ & TiO$_2$,  Ti$_3$O$_5$, CO$_2$,
  & TiO$_2$,  Ti$_3$O$_5$, N$_2$ & TiO$_2$,  Ti$_3$O$_5$, AlF$_3$ &
  -8.37 \\
  C & $-$ &  Ti$_3$O$_5$, CO$_2$,  CO& $-$  & $-$  & -9.24\\
D &   Ti$_3$O$_5$, Ti$_2$O$_3$ &   Ti$_3$O$_5$, Ti$_2$O$_3$, CO,
  &  Ti$_3$O$_5$, Ti$_2$O$_3$, N$_2$ &  Ti$_3$O$_5$,Ti$_2$O$_3$, AlF$_3$ &
  -9.75 \\
  E & $-$ & $-$ & Ti$_2$O$_3$, N$_2$, TiN & $-$  & -9.88 \\
  F & $-$ & Ti$_2$O$_3$,  CO, TiC&$-$ & $-$  & -10.06 \\
  G & $-$ & CO, TiC, C & $-$ & $-$  & -10.24 \\
  H & $-$ & $-$ & N$_2$, TiN, AlN & $-$  & -10.30 \\
  I & Ti$_2$O$_3$, TiO & Ti$_2$O$_3$, TiO, TiC & Ti$_2$O$_3$, TiO, TiN & Ti$_2$O$_3$, TiO, AlF$_3$ & -10.36 \\
 J &  TiO, Al & TiO, Al, TiC & TiO, Al, TiN &TiO, Al, AlF$_3$ & -10.45 \\
    K & $-$ & $-$ & Al, TiN, AlN & $-$ & -10.45 \\
    L & $-$ & Al, TiC, C & $-$ & $-$  & -10.45 \\
  \hline
\end{tabular}

      \end{center}
      \end{table}

Chemical potentials of atoms are calculated for all reference points using the equations
that relate these potentials with the potentials
 Eq. (\ref{18-0}). For instance, for the point A of the ternary
system we have
\begin{eqnarray} \label{20}
  2\mu_{\mathrm{Al}}+3\mu_{\mathrm{O}} &=& \mu(\mathrm{Al}_2\mathrm{O}_3), \\
  \label{21}
  \mu_{\mathrm{Ti}}+2 \mu_{\mathrm{O}}  &=& \mu(\mathrm{Ti}\mathrm{O}_2), \\
  \label{22}
   2\mu_{\mathrm{O}}&=& \mu(\mathrm{O}_2).
\end{eqnarray}

The expression for the defect formation energies  (\ref{10}) contains the difference of
the energies of a defect and of the perfect supercell
$E_{\mathrm{d-p},i}=E_{\mathrm{def},i}-E_{perf}$. Within the DFT method we calculate this
difference at $T=0$.
 At finite $T$ one should take into account a temperature correction $\Delta E_{\mathrm{d-p},i}(T)$
  to this difference.  The origin
 of  $\Delta E_{\mathrm{d-p},i}(T)$ is differences of
 finite temperature enthalpy and  entropy of a defect
and a perfect supercell. We evaluate $\Delta E_{\mathrm{d-p},i}(T)$
from the temperature corrections Eq. (\ref{18}) for relevant
materials in the crystal state.

The difference $\Delta E_{\mathrm{d-p},\mathrm{Ti}_\mathrm{Al}}(T)$ is evaluated from
$\Delta\mu_T$ for pure $\mathrm{Al}_2\mathrm{O}_3$ and $\mathrm{Ti}_2\mathrm{O}_3$:
\begin{equation}\label{19-5}
\Delta E_{\mathrm{d-p},\mathrm{Ti}_\mathrm{Al}}(T)
=\frac{1}{2}\left[\Delta\mu_T(\mathrm{Ti}_2\mathrm{O}_3)
-\Delta\mu_T({\mathrm{Al}_2\mathrm{O}_3})\right].
\end{equation}
To evaluate the temperature correction for the supercell with a vacancy we imply that
\begin{equation}\label{19-2a}
3\Delta E_{\mathrm{d-p},V_\mathrm{O}}(T)+2\Delta
E_{\mathrm{d-p},V_\mathrm{Al}}(T)=-\Delta\mu_T(\mathrm{Al}_2\mathrm{O}_3).
\end{equation}
Equation (\ref{19-2a}) can be justified as follows. If one removes an integer number of
formula units from the supercell and rearranges atoms one can obtain another perfect
supercell. The formation energy of such a "defect" is equal to zero and the temperature
correction to $E_i$ is equal to zero as well. The temperature correction to $E_i$
contains the  corrections to $\mu_X$ and to $E_{\mathrm{def},i}-E_{perf}$ and they should
cancel each other. Equation (\ref{19-2a}) provides the fulfillment of this condition.
Assuming that $\Delta E_{\mathrm{d-p},V_X}(T)$ (with
$X=\mathrm{Al},\mathrm{Ti},\mathrm{O}$), are the same for $\mathrm{Al}_2\mathrm{O}_3$,
$\mathrm{Ti}_2\mathrm{O}_3$ and $\mathrm{Ti}\mathrm{O}_2$, we calculate $\Delta
E_{\mathrm{d-p},V_\mathrm{Al}}(T)$ and $\Delta E_{\mathrm{d-p},V_\mathrm{O}}(T)$ using
equations
\begin{eqnarray}\label{19-3} 3\Delta E_{\mathrm{d-p},V_\mathrm{O}}(T)+2\Delta
E_{\mathrm{d-p},V_\mathrm{Ti}}(T)=-\Delta\mu_T(\mathrm{Ti}_2\mathrm{O}_3),\\
\label{19-4} 2\Delta E_{\mathrm{d-p}, V_\mathrm{O}}(T)+\Delta
E_{\mathrm{d-p},V_\mathrm{Ti}}(T)=-\Delta\mu_T(\mathrm{Ti}\mathrm{O}_2),
\end{eqnarray}
and Eq. (\ref{19-2a}).

Temperature corrections to the energies of substitutional N, C, and F defects are
evaluated from $\Delta\mu_T(\mathrm{AlN})$, $\Delta\mu_T(\mathrm{Al}_4\mathrm{C}_3)$ and
$\Delta\mu_T(\mathrm{Al}\mathrm{F}_3)$, correspondingly. For instance, to evaluate
$\Delta E_{\mathrm{d-p},\mathrm{N}_\mathrm{O}}(T)$ we use the relation
\begin{equation}
  \label{19-7-1}
\Delta E_{\mathrm{d-p},\mathrm{N}_\mathrm{O}}(T) = \Delta\mu_T(\mathrm{Al}\mathrm{N})+
  \Delta E_{\mathrm{d-p},V_\mathrm{Al}}(T)+ \Delta
  E_{\mathrm{d-p},V_\mathrm{O}}(T).
\end{equation}
Equation (\ref{19-7-1}) is based on the fact that the $\mathrm{N}_\mathrm{O}$ defect can
be created by removing of one Al and one O atom from the $\mathrm{Al}_2\mathrm{O}_3$
crystal and adding one formula unit of $\mathrm{Al}\mathrm{N}$. Similar arguments were
used to evaluate $\Delta E_{\mathrm{d-p},\mathrm{C}_\mathrm{O}}(T)$, $\Delta
E_{\mathrm{d-p},\mathrm{C}_\mathrm{Al}}(T)$, and
$E_{\mathrm{d-p},\mathrm{F}_\mathrm{O}}(T)$. The temperature correction $\Delta
E_{\mathrm{d-p,i_c}}(T)$ for a complex defect is taken as the sum of $\Delta
E_{\mathrm{d-p},i}(T)$ for simple defects which form this complex defect.

The correction $E^{(c)}_i$ in the expression for the  defect formation energy (\ref{10})
is evaluated by a method similar to the one proposed in \cite{h09,h10}. DFT supercell
calculations give the formation energies  $E_i^{(0)}$ with spurious electrostatic
interaction between periodically arranged charged defects. Calculations of the charge
distribution show that the defect charge is strongly localized. Therefore, the main
contribution to the electrostatic energy comes from thr monopole-monopole interaction,
while the contribution of the monopole-dipole and dipole-dipole interaction \cite{mp,
komsa} is much smaller. This allows us to equate $E_i^{(c)}$ with the Madelung energy
taken with a negative sign. We imply that the Madelung energy for a supercell composed of
$n\times m\times k$ unit cells can be presented in the form
\begin{equation}\label{20-1}
    E_M(n,m,k)=\frac{q_{i,eff}^2}{2\epsilon_{eff}} \tilde{E}_{M}(n,m,k),
\end{equation}
where  $q_{eff}$ is the effective charge of the defect, $\varepsilon_{eff}$ is the
effective dielectric constant, and $\tilde{E}_{M}(n,m,k)$ is the electrostatic energy of
a periodic structure of $q=1$ point charges in the medium with $\varepsilon=1$ and with a
charge compensating background.
 For a cubic lattice $\tilde{E}_{M}(n,n,n)= -2.837/n a$, where $a$ is the
lattice parameter. In the general case $\tilde{E}_{M}(n,m,k)$ depends on the size and the
form of the supercell. For elongated supercells it can be positive.

\begin{figure}[h]
     \center{\includegraphics[width=0.4\textwidth]{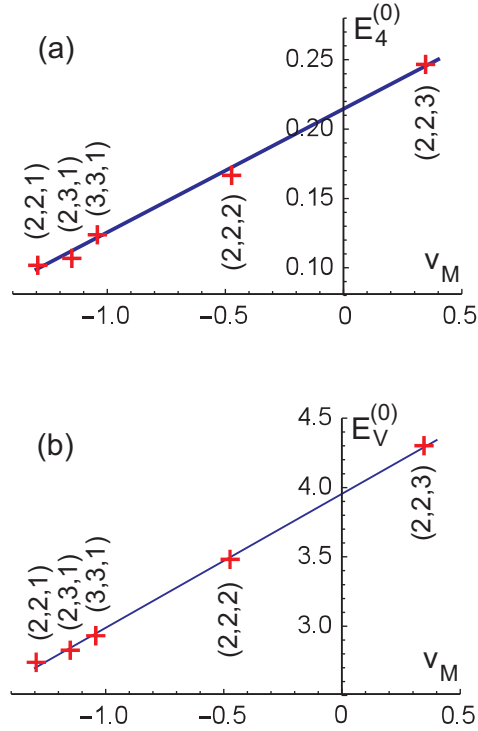}}
     \caption{Calculated  defect formation energies $E^{(0)}_i$ (in eV) for $\mathrm{Ti}^{4+}$ (a)
     and
     $V_\mathrm{Al}^{3-}$ (b) for
     several supercells versus $v_M$ and linear fit (\ref{20-2}).
     For concreteness, we set $\mu_e$ equals to the
     valence band maximum.}\label{f2}
     \end{figure}

For each defect specie we calculate the supercell defect formation energy $E_i^{(0)}$ for
five different supercells: $(2,2,1)$, $(2,3,1)$, $(3,3,1)$, $(2,2,2)$ and $(2,2,3)$. Each
supercell is built from hexagonal unit cells with lattice parameters $a=b=4.86$ $\AA$ and
$c=13.19$ $\AA$ obtained for pure $\mathrm{Al}_2\mathrm{O}_3$ by optimization of lattice
vectors. Then, we calculate the Madelung energy $\tilde{E}_M$ by the Ewalds method for
the same supercells,  define the dimensionless quantity $v_M(n,m,k)=a\tilde{E}_M(n,m,k)$,
and fit the energy $E_i^{(0)}$ by a linear function:
\begin{equation}\label{20-2}
E_i^{(0)}(v_M)=E_i + \frac{q_{i,eff}^2}{2 \epsilon_{eff} a} v_{M}=c_{0,i}+c_{1,i} v_{M}.
 \end{equation}
As two examples we present the result of fitting for $\mathrm{Ti}^{4+}$ and
$V_\mathrm{Al}^{3-}$ in Fig \ref{f2}. Taking into account that for infinitely large
supercell $v_{M}(n,m,k)= 0$, we evaluate the corrected formation energy as $E_i=c_{0,i}$
and the correction $E_i^{(c)}$, as $E_i^{(c)}(n,m,k)=c_{0,i}-E_i^{(0)}(n,m,k)$.
Calculated electrostatic corrections $E_i^{(c)}(2,2,1)$ for the considered charged
defects are given in Table \ref{t2}. Note that if we set $\varepsilon_{eff}=10$ we obtain
the effective charges $|q_{\mathrm{Ti}^{4+},eff}/e|=0.76$ and
$|q_{V_\mathrm{Al}^{3-},eff}/e|=2.52$ which looks reasonable.

\begin{table}
  \centering
  \caption{Electrostatic correction $E_i^{(c)}$ to the defect formation energy for
  the $2\times 2\times 1$ supercell}\label{t2}
  \begin{tabular}{|r|r|r|}
    \hline
    Defect & q &  $E_i^{(c)}$, eV \\ \hline
    $\mathrm{Ti}^{4+}$ & $+1$ &0.11 \\
    $\mathrm{V}_\mathrm{Al}^{3-}$ & $-3$ & 1.21 \\
    $\mathrm{Ti}^{4+}$-$\mathrm{V}_\mathrm{Al}^{3-}$ & $-2$ & 0.42 \\
    $\mathrm{Ti}^{3+}$-$\mathrm{Ti}^{4+}$ & $+1$& 0.05 \\
     $\mathrm{Ti}^{4+}$-$\mathrm{Ti}^{4+}$-$\mathrm{V}_\mathrm{Al}^{3-}$ &$-1$&
     0.21 \\
    $\mathrm{Ti}^{3+}$-$\mathrm{Ti}^{4+}$-$\mathrm{V}_\mathrm{Al}^{3-}$ & $-2$& 0.68 \\
$\mathrm{Ti}^{3+}$-$\mathrm{Ti}^{3+}$-$\mathrm{V}_\mathrm{Al}^{3-}$ & $-3$& 1.48 \\
    $\mathrm{F}_\mathrm{O}^+$ &  +1& 0.14 \\
    $\mathrm{F}_\mathrm{O}^+$-$\mathrm{V}_\mathrm{Al}^{3-}$ &  $-2$& 0.54 \\
   $\mathrm{F}_\mathrm{O}^+$-$\mathrm{F}_\mathrm{O}^+$-$\mathrm{V}_\mathrm{Al}^{3-}$ &  $-1$& 0.16 \\
    $\mathrm{N}_\mathrm{O}^-$& $-1$ & 0.17 \\
    $\mathrm{C}_\mathrm{O}^{2-}$& $-2$ & 0.60 \\
    $\mathrm{C}_\mathrm{O}^{-}$& $-1$ & 0.15 \\
    $\mathrm{C}_\mathrm{O}^{2-}$-$\mathrm{Ti}^{4+}$& $-1$ & 0.16 \\
    \hline
  \end{tabular}
\end{table}

\section{Equilibrium concentrations of defects in $\mathbf{Ti}$:sapphire
} \label{s4}

As was already mentioned in the Introduction 
 the
equilibrium concentration of a given charged defect depends on the formation energies of
all charged defects. If one considers the additional constraint (\ref{6}) one should know
the formation energies of all uncharged defects as well. In practice defect species with
large formation energies can be ignored since their contribution in Eqs. (\ref{3}) and
(\ref{6}) is negligible.

We have calculated formation energies of different defects including substitutional and
interstitial ions of Ti, C, N, and F  in different charge states,  native defects (Al and
O vacancies and interstitials), and complexes of such defects. We have found that most of
defects have rather large formation energies.
 We restrict our analysis with
several defects species with the smallest formation energies. In the case of Ti:sapphire
without co-dopants we consider three species of isolated  defects ( $\mathrm{Ti}^{3+}$,
$\mathrm{Ti}^{4+}$ and $V_\mathrm{Al}^{3-}$), three pairs
($\mathrm{Ti}^{3+}$-$\mathrm{Ti}^{3+}$, $\mathrm{Ti}^{3+}$-$\mathrm{Ti}^{4+}$ and
$\mathrm{Ti}^{4+}$-$V^{3-}_\mathrm{Al}$), four triples
($\mathrm{Ti}^{3+}$-$\mathrm{Ti}^{3+}$-$\mathrm{Ti}^{3+}$,
$\mathrm{Ti}^{4+}$-$\mathrm{Ti}^{4+}$-$V^{3-}_\mathrm{Al}$,
$\mathrm{Ti}^{3+}$-$\mathrm{Ti}^{4+}$-$V^{3-}_\mathrm{Al}$ and
$\mathrm{Ti}^{3+}$-$\mathrm{Ti}^{3+}$-$V^{3-}_\mathrm{Al}$), and a quadruple complex
$\mathrm{Ti}^{4+}$-$\mathrm{Ti}^{4+}$-$\mathrm{Ti}^{4+}$-$V^{3-}_\mathrm{Al}$. For
Ti:sapphire with co-dopants we add to this list two or three defect species (see below)
formed by co-dopants.

In this section we concentrate on Ti:sapphire without co-dopants.

If the total number of Ti atoms is not fixed (the constraint (\ref{6}) is not applied)
the concentrations of charged defects are found from the following system of equations:
\begin{eqnarray} \label{cn1}
  \tilde{n}_4^3 \tilde{n}_V = e^{-\frac{3 E_4+E_V}{k_b T}}, \\ \label{cn2}
  \tilde{n}_4 +2 C_{34} \tilde{n}_{3}\tilde{n}_{4}- 3 \tilde{n}_V - 8 C_{4V} \tilde{n}_4 \tilde{n}_V
  - 18 C_{44V}\tilde{n}_4^2 \tilde{n}_V
  -36 C_{34V} \tilde{n}_3\tilde{n}_4 \tilde{n}_V-54  C_{33V}\tilde{n}_3^2 \tilde{n}_V=
  0,
\end{eqnarray}
where we use the notation $V\equiv V^{3+}_\mathrm{Al}$.  In Eq. (\ref{cn2}) we take into
account the relation (\ref{4}). The factors $C_{i_c}$  are the coefficients in the
relation (\ref{4}) averaged over different orientations and configurations:
\begin{equation}\label{p1}
    C_{i_c}=\frac{\sum_f K_{i_c,f} e^{\frac{E_{i_c,f}^{(b)}}{k_B T}}}{\sum_f K_{i_c,f}},
\end{equation}
where $E_{i_c,f}^{(b)}$ is the binding energy of the complex $i_c$ and $K_{i_c,f}$ is the
number of configurations and orientations with the same energy  (different $f$ label
configurations and orientations with distinct energies). Calculated binding energies and
the numbers $K_{i_c,f}$ are given in Table \ref{t3}. Binding energies are independent of
the chemical potentials of atoms. Binding energies given in Table \ref{t3}
 include the electrostatic correction
$E_i^{(c)}$. They differ from ones calculated in Refs. \cite{d2,my}, where this
correction was not taken into account. The concentration of electrically neutral defects
 depends only on its own
formation energy:
\begin{equation}\label{p2}
    \tilde{n}_i=e^{-\frac{E_i}{k_B T}}.
\end{equation}

\begin{table}
  \centering
  \caption{Binding energies $E_{i_c,f}^{(b)}$   and the numbers
  of different orientations and configurations $K_{i_c,f}$ with the same energy.
Complexes of  Al vacancy surrounded by two Ti and Ti surrounded by Ti and Al vacancy are
notated as $\mathrm{Ti}-V-\mathrm{Ti}$ and $\mathrm{Ti}-\mathrm{Ti}-V$, correspondingly.
For complexes $\mathrm{Ti}-\mathrm{Ti}-\mathrm{Ti}-V$ we consider only configurations
with the vacancy in the center.}\label{t3}
\begin{tabular}{|r|c|r|r|}
  \hline
  Complex defect $i_c$ & Notation &$K_{i_c, f}$ & $E_{i_c,f}^{(b)}$, eV \\ \hline
 \multirow{2}{*}{ $\mathrm{Ti}^{3+}$-$\mathrm{Ti}^{3+}$} & \multirow{2}{*}{33}
  &3 & 1.36 \\
   & & 1 & 1.22 \\ &&&\\
  $\mathrm{Ti}^{3+}$-$\mathrm{Ti}^{3+}$-$\mathrm{Ti}^{3+}$
    &333&6 & 2.03 \\ &&&\\
  $\mathrm{Ti}^{4+}$-$\mathrm{Ti}^{4+}$-$\mathrm{Ti}^{4+}$-$V_{\mathrm{Al}}^{3-}$
  &444V &4 & 4.47 \\ &&&\\
  \multirow{2}{*}{ $\mathrm{Ti}^{3+}$-$\mathrm{Ti}^{4+}$}&\multirow{2}{*}{34} & 3 & 0.76 \\
    &&1 & 0.68 \\ &&&\\
 \multirow{2}{*}{  $\mathrm{Ti}^{4+}$-$V_{\mathrm{Al}}^{3-}$ }&\multirow{2}{*}{4V}& 3 & 2.05 \\
    &&1 & 1.85 \\ &&&\\
 \multirow{2}{*}{ $\mathrm{Ti}^{4+}$-$V_\mathrm{Al}^{3-}$-$\mathrm{Ti}^{4+}$}&
 \multirow{4}{*}{44V}   & 3 & 3.36 \\
 & &3& 2.91\\
  \multirow{2}{*}{ $\mathrm{Ti}^{4+}$-$\mathrm{Ti}^{4+}$-$V_\mathrm{Al}^{3-}$ }
   & &6 & 2.51 \\
   && 6 & 2.40 \\ &&&\\
  \multirow{2}{*}{ $\mathrm{Ti}^{3+}$-$V_\mathrm{Al}^{3-}$-$\mathrm{Ti}^{4+}$}
  & \multirow{4}{*}{34V}  & 3 & 2.69 \\
  & & 3 & 2.51 \\
 \multirow{2}{*}{$\mathrm{Ti}^{3+}$-$\mathrm{Ti}^{4+}$-$V_\mathrm{Al}^{3-}$ }
 & & 6 & 2.55 \\
  & & 6 & 2.49 \\ &&&\\
{ $\mathrm{Ti}^{3+}$-$V_\mathrm{Al}^{3-}$-$\mathrm{Ti}^{3+}$ }&
 \multirow{2}{*}{33V}  &6 &
0.92 \\
   $\mathrm{Ti}^{3+}$-$\mathrm{Ti}^{3+}$-$V_\mathrm{Al}^{3-}$ && 12 & 1.88
   \\&&& \\
\multirow{2}{*}{ $\mathrm{N}_\mathrm{O}^-$-$\mathrm{Ti}^{4+}$} &\multirow{2}{*}{-}& 3 & 1.35 \\
&&3&1.10\\&&&
 \\
 { $\mathrm{C}_\mathrm{O}^{2-}$-$\mathrm{Ti}^{4+}$} &-& 6 & 2.95
 \\&&&\\
 \multirow{2}{*}{ $\mathrm{F}_\mathrm{O}^+$-$V_\mathrm{Al}^{3-}$}&\multirow{2}{*}{FV} & 3 & 1.61
 \\ &&3&2.03\\&&&
 \\
 \multirow{4}{*}{  $\mathrm{F}_\mathrm{O}^+$-$V_\mathrm{Al}^{3-}$-$\mathrm{F}_\mathrm{O}^+$ }
 &\multirow{4}{*}{FVF}& 3 & 3.44 \\
   && 3 & 2.87 \\
   && 3 & 3.79 \\
   && 6 & 3.30 \\
    \hline
\end{tabular}
\end{table}

The formation energies of $\mathrm{Ti}^{3+}$ and of the electrically neutral combination
of three $\mathrm{Ti}^{4+}$ and one $V_\mathrm{Al}^{3-}$ at $T=T_m$ are given in Table
\ref{t4} for all reference points of the $\mathrm{Al}-\mathrm{O}-\mathrm{Ti}$ phase
diagram. Using Eqs. (\ref{cn1}), (\ref{cn2}) and (\ref{p2}) and the data from Tables
\ref{t3} and \ref{t4} we calculate $\tilde{n}_3$, $\tilde{n}_4$ and $\tilde{n}_V$.
Concentrations of  complex defects are found from Eq. (\ref{4}) in which the factor
$\exp(E^{(b)}_{i_c}/k_B T)$ is replaced with the coefficient $C_{i_c}$ Eq. (\ref{p1}).
The concentration of an $i$-th defect specie is calculated as $c_i=K_i \tilde{n}_i
c_{\mathrm{Al}}$, where $c_{\mathrm{Al}}=4.7\cdot 10^{22}$ cm$^{-3}$ is the concentration
of Al in the $\mathrm{Al}_2\mathrm{O}_3$ crystal, and $K_i=\sum_fK_{i,f}$ is the total
number of configurations and orientations of  the $i$-th defect specie ($K_i=1$ for
$\mathrm{Ti}^{3+}$, $\mathrm{Ti}^{4+}$ and $V_\mathrm{Al}^{3-}$).

\begin{table}
  \centering
  \caption{Defect formation energies (in eV) at reference points
  of $\mathrm{Al}-\mathrm{O}-\mathrm{Ti}$
  and $\mathrm{Al}-\mathrm{O}-\mathrm{Ti}-X$ ($X=\mathrm{C}, \mathrm{N}, \mathrm{F}$)
  phase diagram at $T=T_m$.  The energies of electrically neutral combinations
   are given per one defect.}
  \label{t4}
\begin{tabular}{|c|c|c|c|c|c|c|}
  \hline
  Point & $E_3$ & $(3E_4+E_V)/4$ & $(E_4+E_{\mathrm{N}^-_\mathrm{O}})/2$ &
  $(2E_4+E_{\mathrm{C}_\mathrm{O}^{2-}})/3$ &$(E_4+E_{\mathrm{C}_\mathrm{O}^{-}})/2$
  &$E_{\mathrm{F}_\mathrm{O}^+}-E_4$ \\ \hline
  A & 2.25 & 1.57 & 3.21 & 5.64 &7.24&0.29\\
  B & 1.71 & 1.57 & 2.40 & 4.19 &5.33&0.29\\
  C & 1.57  & 1.79 & - & 3.23 &3.97&-\\
  D & 1.48  & 1.92 & 1.59 & 2.83 &3.41&$-$0.17\\
  E & 1.48 & 1.96 & 1.53 & - &&-\\
  F & 1.48 & 2.03 & - & 2.62 &3.09&-\\
  G & 1.92 & 2.43 & - & 2.80 &3.14&-\\
  H & 2.11 & 2.59 & 1.64 & - &&-\\
  I & 1.48 & 2.14 & 1.65 & 2.67 &3.17&$-$0.47\\
  J & 1.52 & 2.21 & 1.68 & 2.70 &3.19&$-$0.56\\
  K & 2.11 & 2.65 & 1.68 & - &&-\\
  L & 2.23 & 2.74 & - & 2.94 &3.19&-\\
  \hline
\end{tabular}
\end{table}

\begin{figure}
\includegraphics[width=0.4\linewidth]{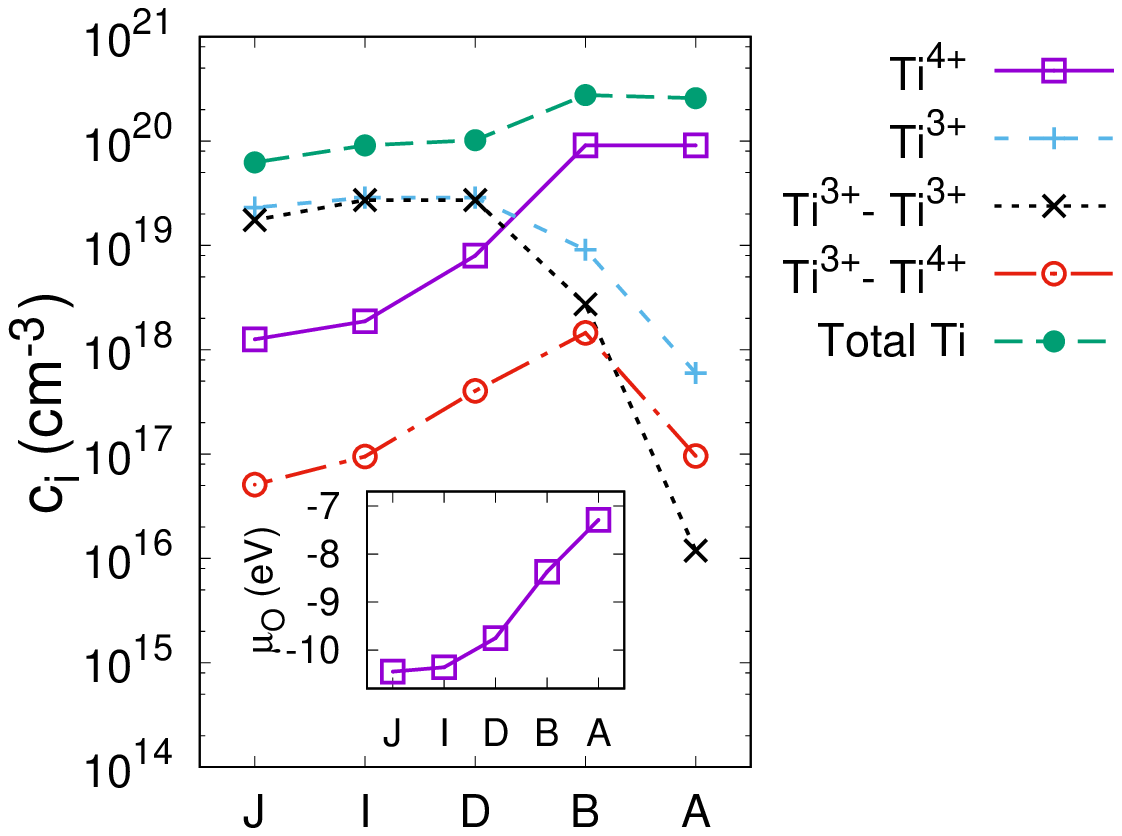}
\includegraphics[width=0.47\linewidth]{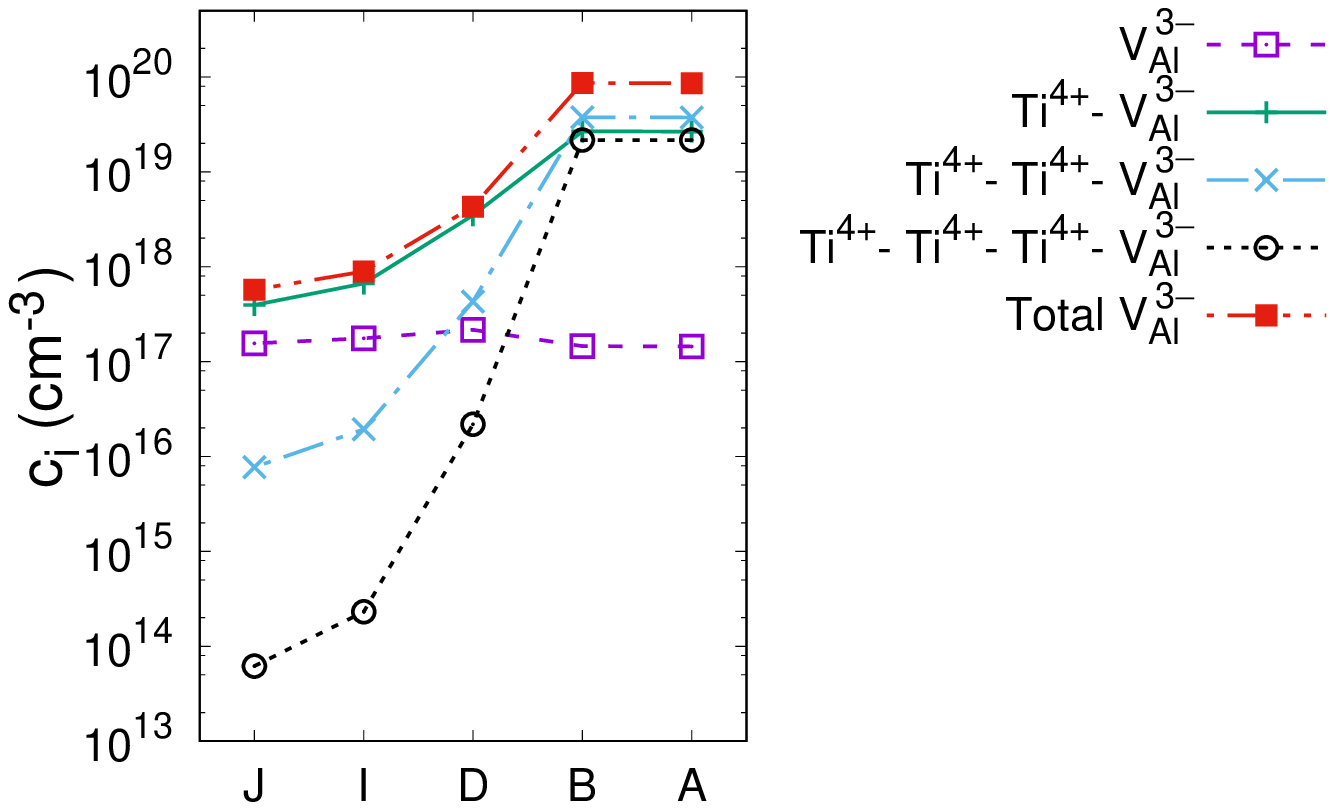}
\caption{Equilibrium concentrations of defects in Ti:sapphire at $T=T_m$ in the
conditions that correspond to reference points of $\mathrm{Al}-\mathrm{Ti}-\mathrm{O}$
phase diagram Fig. \ref{f1} (see also Table \ref{t1}). Lines are guides to the eye.
Oxygen chemical potential is given in the inset.} \label{f3}
\end{figure}

The obtained equilibrium concentrations of defects and the overall concentration of Ti
are shown in Fig. \ref{f3}.  To make this figure more readable we do not display the
concentration of $\mathrm{Ti}^{3+}$-$\mathrm{Ti}^{3+}$-$\mathrm{Ti}^{3+}$ triples which
is lower than that of $\mathrm{Ti}^{3+}$-$\mathrm{Ti}^{3+}$ pairs. The concentrations of
the
 $\mathrm{Ti}^{3+}$-$\mathrm{Ti}^{3+}$-$V^{3-}_\mathrm{Al}$,
and $\mathrm{Ti}^{3+}$-$\mathrm{Ti}^{4+}$-$V^{3-}_\mathrm{Al}$ complexes are not
displayed either. They are generally much lower than the concentration of
$\mathrm{Ti}^{4+}$-$\mathrm{Ti}^{4+}$-$V^{3-}_\mathrm{Al}$ complexes (they are comparable
only in the reduced conditions).

One can see from Fig. \ref{f3} that the overall concentration of Ti varies in the range
from $6\cdot 10^{19}$ cm$^{-3}$ to $3.5\cdot 10^{20}$ cm$^{-3}$ that corresponds to the
range  from $0.12$ wt\% to 0.7 wt\% of Ti. This concentration is rather high. For
instance,  in samples investigated in \cite{m1}  the concentration of Ti was in the range
from 0.006 to 0.2 wt\%. Samples used in Refs.\cite{1,2,uv,3a,pinto} had the concentration
of Ti less than 0.1 wt\%. The concentration of Ti  in samples grown in \cite{nizh} was in
the range from 0.1 to 0.25 wt\%.

If the total concentration of Ti differs from the equilibrium one we should take into
account the constraint (\ref{6}). Then we arrive at a system of three equations:
\begin{eqnarray}
\label{101}
  \tilde{n}_4^3 \tilde{n}_V = \tilde{n}_3^3e^{\frac{3 E_3-3 E_4-E_V}{k_b T}}, \\
  \label{101a}
\tilde{n}_4 +2 C_{34} \tilde{n}_{3}\tilde{n}_{4}- 3 \tilde{n}_V - 8 C_{4V} \tilde{n}_4
\tilde{n}_V
  - 18 C_{44V}\tilde{n}_4^2 \tilde{n}_V
  -36 C_{34V} \tilde{n}_3\tilde{n}_4 \tilde{n}_V-54  C_{33V}\tilde{n}_3^2 \tilde{n}_V=
  0,\\
  \label{102}
   \tilde{n}_3 + \tilde{n}_4 +4 C_{34} \tilde{n}_{3}\tilde{n}_{4} +4 C_{33} \tilde{n}_{3}^2
 +18 C_{333} \tilde{n}_{3}^3  +  4 C_{4V} \tilde{n}_4 \tilde{n}_V+
   36 C_{44V}\tilde{n}_4^2 \tilde{n}_V \cr
  +36 C_{34V} \tilde{n}_3\tilde{n}_4 \tilde{n}_V+ 36  C_{33V}\tilde{n}_3^2
  \tilde{n}_V+
  12 C_{444V} \tilde{n}_4^3 \tilde{n}_V= \frac{n_{\mathrm{Ti}}}{n_{\mathrm{Al}}}.
\end{eqnarray}

.

The quantity $3E_3-3 E_4-E_V$  in Eq. (\ref{101}) and the coefficients $C_{i_c}$ in Eqs.
(\ref{101a}), and (\ref{102}) are independent of $\mu_\mathrm{Ti}$. Consequently, the
solution of the system (\ref{101}), (\ref{101a}), and (\ref{102}) is also independent of
$\mu_\mathrm{Ti}$ and concentrations can be presented as continuous functions of
$\mu_\mathrm{O}$. The oxygen chemical potential varies in the range
$[\mu_{\mathrm{r}},\mu_{\mathrm{o}}]$, where
$\mu_{\mathrm{r}}=[\mu(\mathrm{Al}_2\mathrm{O}_3)-2\mu(\mathrm{Al})]/3$
($\mu_{\mathrm{o}}=\mu(\mathrm{O}_2)/2$) is the value of the oxygen chemical potential in
the reduced (oxidized) limit. The boundaries $\mu_{\mathrm{r}}$ and $\mu_{\mathrm{o}}$
depend on temperature.

\begin{figure}
\includegraphics[width=0.5\linewidth]{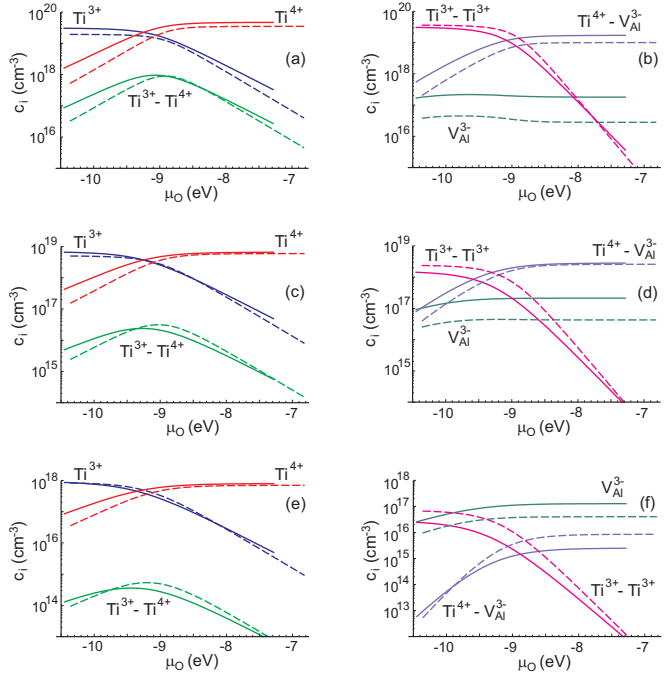}
\caption{Equilibrium concentrations of isolated and paired defects at $T=2327$ K (solid
curves) and $T=2000$ K (dashed curves) at fixed overall concentration of Ti
$c_\mathrm{Ti}=10^{20}$ cm$^{-3}$ (a,b), $c_\mathrm{Ti}=10^{19}$ cm$^{-3}$ (c,d) and
$c_\mathrm{Ti}=10^{18}$ cm$^{-3}$ (e,f).} \label{f4}
\end{figure}

Calculated concentrations of simple and paired defects at fixed overall concentration of
Ti ($c_{\mathrm{Ti}}$) are presented in Fig. \ref{f4}. We consider three different
$c_\mathrm{Ti}$ (10$^{20}$ cm$^{-3}$, $10^{19}$ cm$^{-3}$ and $10^{18}$ cm$^{-3}$) and
two temperatures:  $T=T_m=2327$ K and  $T=T_a=2000$ K. We consider that $T=T_m$
concentrations correspond to as-grown samples and $T=T_a<T_m$ ones,  to samples annealed
at $T=T_a$. One can see from Fig. \ref{f5} that annealing may result in an increase or a
decrease of the concentration of a given defect species depending on the oxygen chemical
potential and the total concentration of Ti. As was expected,  a relative fraction of
isolated defects increases under lowering in the total concentration of Ti.

\section{Concentration of defects in Ti:sapphire grown in the presence
of carbon, nitrogen, fluorine and their compounds} \label{s5}

 To obtain crystals with a low
concentration of $\mathrm{Ti}^{4+}$ and high FOM it is desirable to grow them in the
reduced conditions. To do that one can add certain  substances or compounds to the melt
or to the atmosphere or anneal samples in their presence. A side effect of such a
treatment is codoping of Ti:sapphire. Charged defects formed by dopants may change a
balance between  different Ti defects which results in a change of the ratio of
concentration of isolated Ti$^{3+}$ ions to the concentration of Ti$^{3+}-$Ti$^{4+}$
pairs. The mechanism is quite simple. If codoping results in an appearance of additional
positively charged defects, some negatively charged vacancies $V_{\mathrm{Al}}^{3-}$
compensate codopants and the concentration of $\mathrm{Ti}^{4+}$ decreases. Consequently,
according to Eq. (\ref{13}) (valid for codoped samples, as well) the ratio
$c_{\mathrm{Ti}^{3+}}/c_{\mathrm{Ti}^{3+}-\mathrm{Ti}^{4+}}$ increases. If codoping
results in the appearance of additional negatively charged defects, the concentration of
isolated $\mathrm{Ti}^{4+}$ ions increases to provide charge compensation. Consequently,
the ratio $c_{\mathrm{Ti}^{3+}}/c_{\mathrm{Ti}^{3+}-\mathrm{Ti}^{4+}}$ decreases. To
determine the role of a given co-dopant one should take into account not only simple
charged defects formed by co-dopant atoms but complex defects as well. Among such complex
defects it is important to consider pairs and triples formed by co-dopants with vacancies
$V_{\mathrm{Al}}^{3-}$ and with $\mathrm{Ti}^{4+}$.

In addition to a shift of the charge balance, co-doping may create additional impurity
levels in the band gap that will influence optical properties of crystals.

In this section we consider co-doping of Ti:sapphire with carbon, nitrogen and fluorine.

\subsection{$\mathrm{Al}-\mathrm{Ti}-\mathrm{O}-\mathrm{C}$ system}

\begin{figure}
\includegraphics[width=0.5\linewidth]{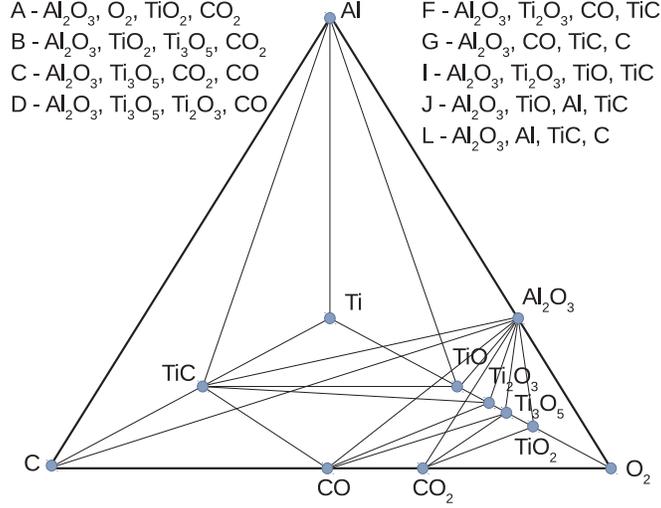}
\caption{The phase diagram of the Al$-$Ti$-$O$-$C system at $T=T_m$ and reference
points.} \label{f5}
\end{figure}

Using calculated heats of formation of relevant materials we build the four-component
$\mathrm{Al}-\mathrm{Ti}-\mathrm{O}-\mathrm{C}$ phase diagram
 (Fig. \ref{f5}). The calculations were done for graphite phase
of carbon.
There are nine points at the diagram Fig. \ref{f5} (labeled as A, B, C, D, F, G, I, J,
L), where $\mathrm{Al}_2\mathrm{O}_3$ is in equilibrium with three other phases.

We have calculated the formation energies of substitutional $\mathrm{C}_\mathrm{O}$ and
$\mathrm{C}_\mathrm{Al}$ defects and carbon interstitial in different charge states. Most
of these energies are quite large. In the intermediate and reduced conditions two
substitutional defects, $\mathrm{C}_\mathrm{O}^{2-}$ and $\mathrm{C}_\mathrm{O}^{-}$,
dominate. Negatively charged carbon defects may bind in pairs and triples with
$\mathrm{Ti}^{4+}$ and $\mathrm{Ti}^{3+}$. Binding energies of such complexes are large
and complex defects prevail over isolated ones. Among complex defects the concentration
of $\mathrm{C}_\mathrm{O}^{2-}$-$\mathrm{Ti}^{4+}$ pairs is the highest one. It can reach
the level of
 $10^{16}$ cm$^{-3}$. But since the concentration of isolated $\mathrm{Ti}^{4+}$ is at
least two orders  higher, carbon defects practically do not influence the concentration
of $\mathrm{Ti}^{4+}$ and $\mathrm{Ti}^{3+}$-$\mathrm{Ti}^{4+}$.

Our results correlate with ones obtained for C-doped $\alpha$-Al$_2$O$_3$ in
\cite{cn1,c2,c3}, where the formation energies of substitutional and interstitial carbon
defects were calculated. It was shown in \cite{cn1,c2,c3} that in the reduced conditions
the substitutional $\mathrm{C}_{\mathrm{O}}$ defects have the smallest formation energy,
while in the oxidized conditions the substitutional $\mathrm{C}_{\mathrm{Al}}$ defects
are energetically preferable. The formation energies of interstitial carbon defects are
large both in the reduced and in the oxidized conditions. Here we do not consider
$\mathrm{C}_{\mathrm{Al}}$ but find that in the oxidized conditions the concentrations of
$\mathrm{C}_{\mathrm{O}}$ are extremely low ones (see Fig. \ref{f7} below). We cannot
compare directly the results of \cite{cn1,c2,c3} with our results since in
\cite{cn1,c2,c3} the chemical potential of C was set to be the same as in diamond
\cite{cn1} or in graphite \cite{c2,c3} irrespective of the value of the oxygen chemical
potential.

\begin{figure}
\includegraphics[width=0.4\linewidth]{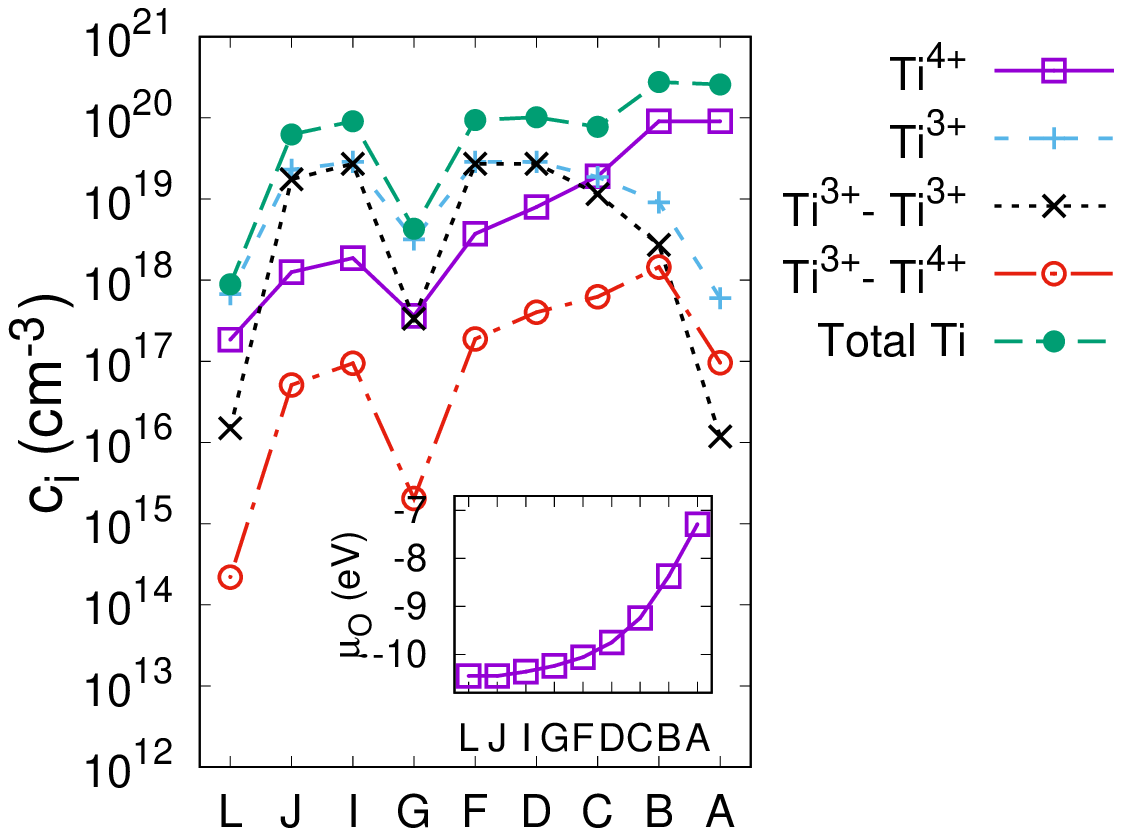}
\includegraphics[width=0.47\linewidth]{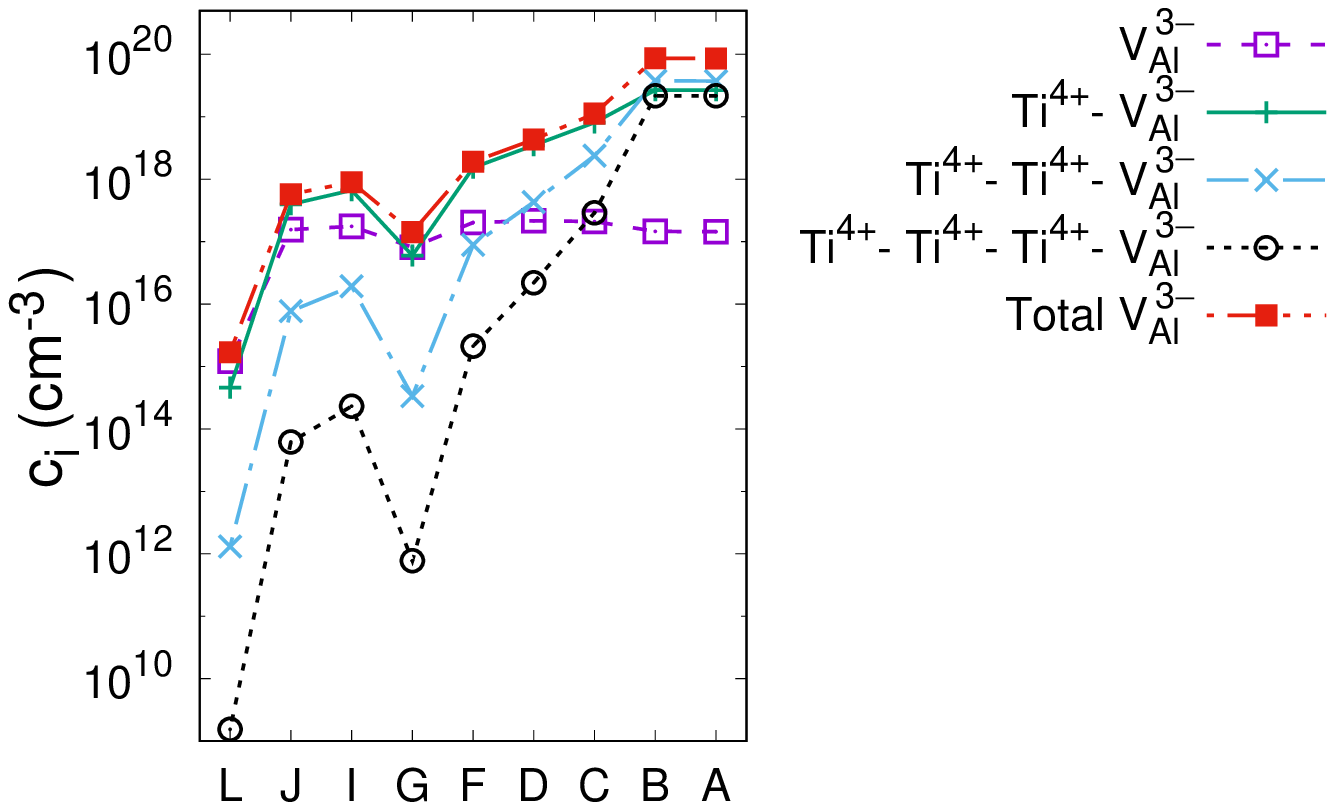}
\caption{Equilibrium concentrations of defects in Ti:sapphire at $T=T_m$  in the
conditions that correspond to reference points of
$\mathrm{Al}-\mathrm{Ti}-\mathrm{O}-\mathrm{C}$ phase diagram Fig. \ref{f5}} \label{f6}
\end{figure}

In Fig.  \ref{f6} we display equilibrium concentrations of Ti defects and Al vacancies at
the reference points of the Al$-$Ti$-$O$-$C phase diagram at $T=T_m$. The concentrations
are obtained from Eqs. (\ref{cn1}) and (\ref{cn2}), where we neglect the contribution of
carbon defects.

The concentrations of carbon defects are found from the relations
\begin{equation}\label{cc2}
    \tilde{n}_{\mathrm{C}_\mathrm{O}^{-}}=\frac{e^{-\frac{E_{\mathrm{C}_\mathrm{O}^{-}}+ E_4}{k_B
    T}}}{\tilde{n}_4},\quad \tilde{n}_{\mathrm{C}_\mathrm{O}^{2-}}=
    \frac{e^{-\frac{E_{\mathrm{C}_\mathrm{O}^{2-}}+
    2 E_4}{k_B
    T}}}{(\tilde{n}_4)^2}.
\end{equation}
Equations (\ref{cc2}) are obtained from Eq. (\ref{7}) (with $\lambda_\mathrm{Ti}=0$).
The concentration of pairs $\mathrm{C}_\mathrm{O}^{2-}-\mathrm{Ti}^{4+}$ is found from
Eq. (\ref{4}).
 The energies $(E_{\mathrm{C}_\mathrm{O}^{-}}+E_4)/2$ and $(E_{\mathrm{C}_\mathrm{O}^{-}}+2E_4)/3$
 are given in Table
\ref{t4}, and the binding energy of the $\mathrm{C}_\mathrm{O}^{2-}-\mathrm{Ti}^{4+}$
pair, in Table \ref{t3}. Calculated concentrations of carbon defects are shown in Fig.
\ref{f7}.

\begin{figure}
\includegraphics[width=0.4\linewidth]{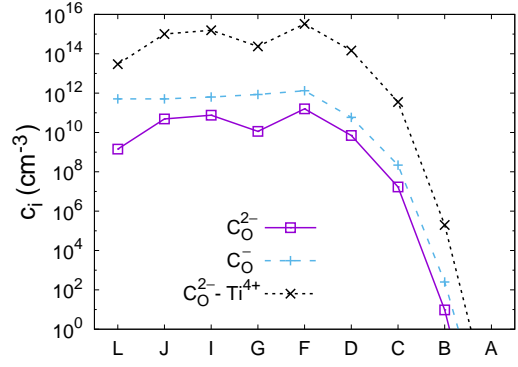}
\caption{ Equilibrium concentrations of carbon defects at $T=T_m$. Lines are guides to
the eye.} \label{f7}
\end{figure}

\subsection{$\mathrm{Al}-\mathrm{Ti}-\mathrm{O}-\mathrm{N}$ system}

\begin{figure}
\includegraphics[width=0.5\linewidth]{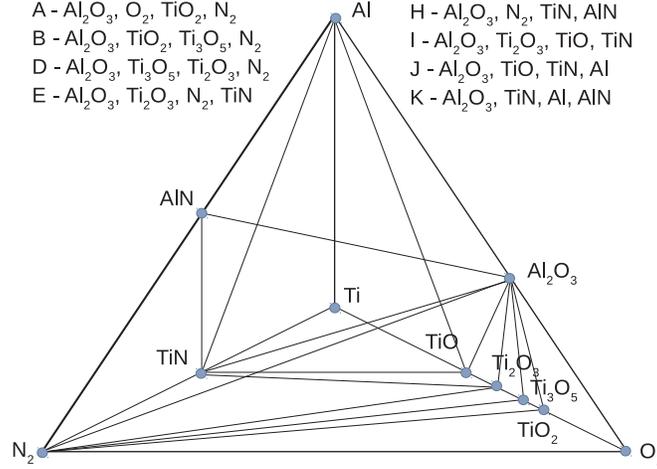}
\caption{The phase diagram of the Al$-$Ti$-$O$-$N system at $T=T_m$ and reference
points.} \label{f8}
\end{figure}

\begin{figure}
\includegraphics[width=0.4\linewidth]{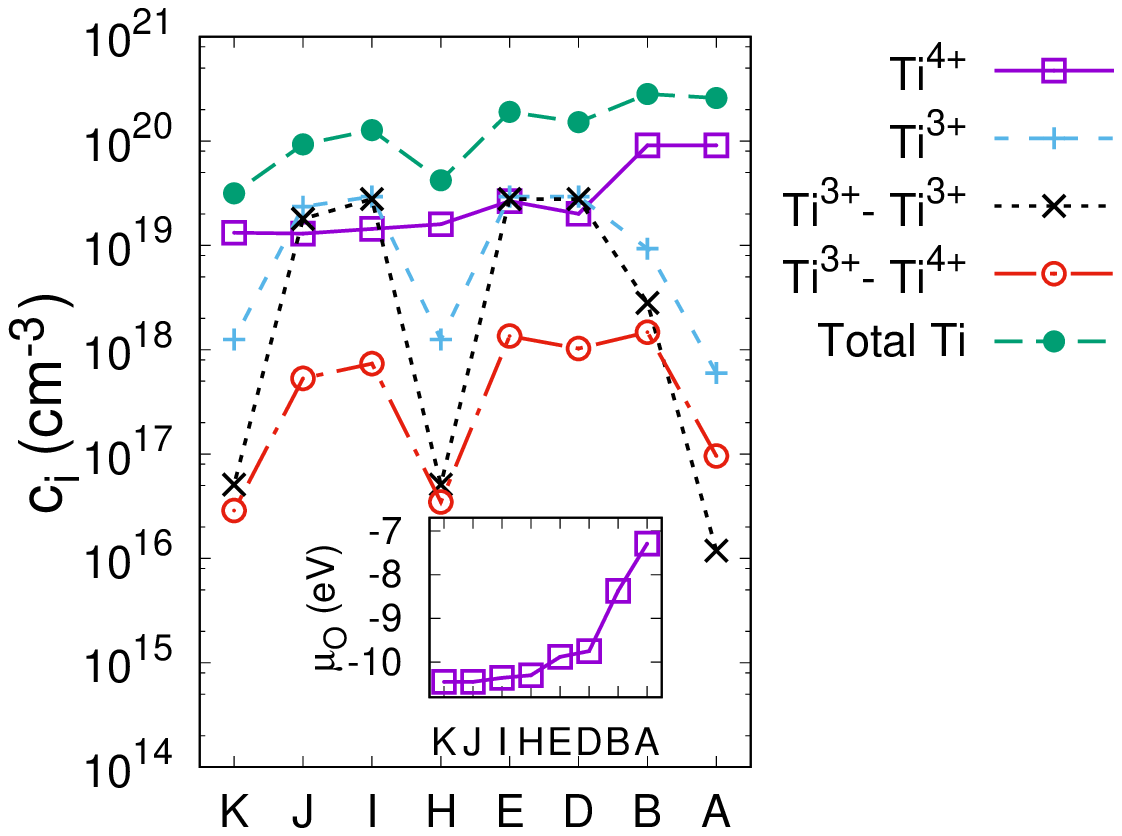}
\includegraphics[width=0.47\linewidth]{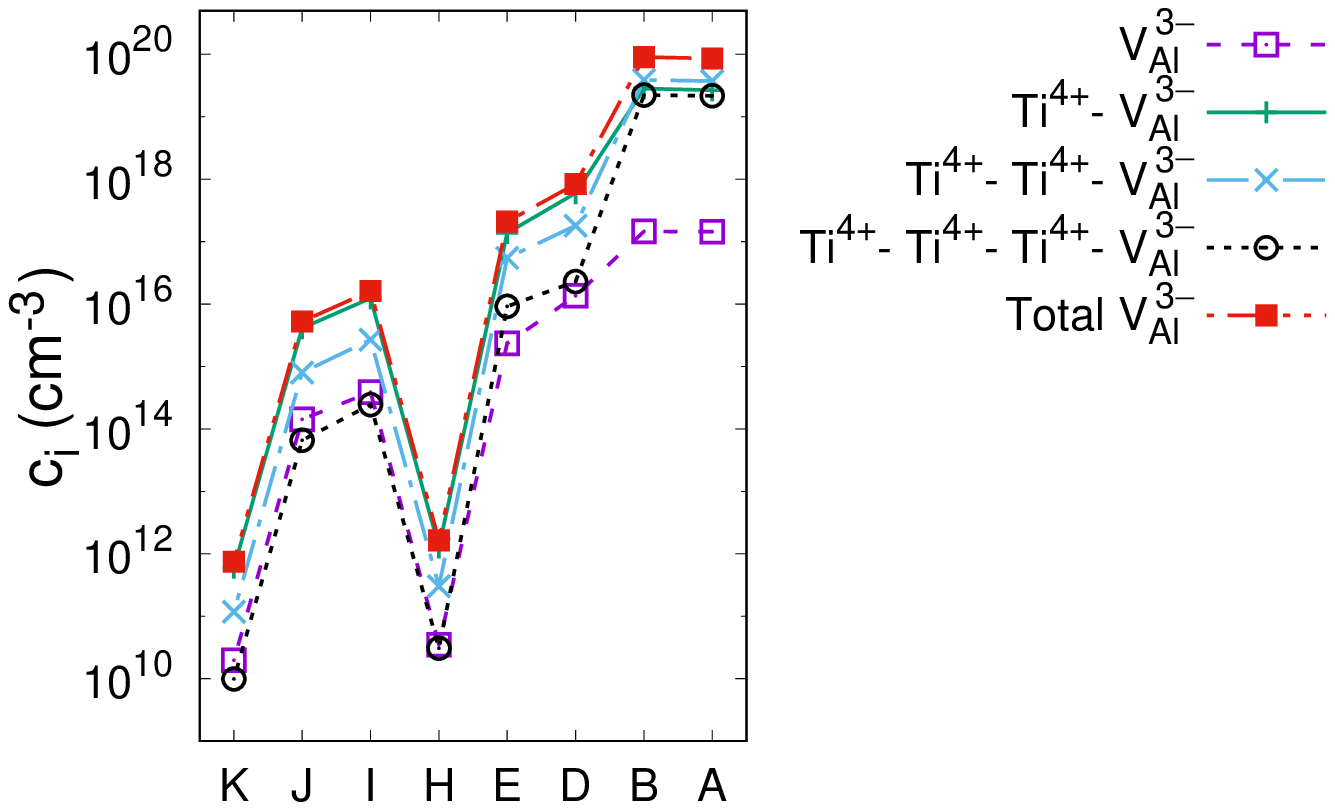}
\caption{ Equilibrium concentrations of defects in Ti:sapphire at $T=T_m$ in the
conditions that correspond to reference points of
$\mathrm{Al}-\mathrm{Ti}-\mathrm{O}-\mathrm{N}$ phase diagram Fig. \ref{f8}.} \label{f9}
\end{figure}

In Fig. \ref{f8} we present the  $\mathrm{Al}-\mathrm{Ti}-\mathrm{O}-\mathrm{N}$ phase
diagram at $T=T_m$ that was built using calculated heats of formation of related
materials. This diagram contains eight reference points (A, B, D, E, H, I, J, K).
The formation energy of $\mathrm{N}_\mathrm{O}^-$ defects is small and their
concentration is high. It correlates with the results of \cite{cn1} where the formation
energies of nitrogen defects in Al$_2$O$_3$ were calculated. It was shown \cite{cn1} that
even in the oxidized conditions the substitutional $\mathrm{N}_{\mathrm{O}}^-$ defect has
the smallest formation energy. The $\mathrm{N}_\mathrm{O}^-$ defects shift the balance
between other charged defects in Ti:sapphire. To evaluate this shift we add the term
$-(3/2) \tilde{n}_{\mathrm{N}}$ to the left-hand side of Eq. (\ref{cn2}), where $
\tilde{n}_\mathrm{N}$, is the number of $\mathrm{N}_\mathrm{O}^-$ defects normalized to
the total number of oxygen atoms.  The quantity $ \tilde{n}_\mathrm{N}$ is expressed
through $\tilde{n}_4$ using Eq. (\ref{7}) with $\lambda_\mathrm{Ti}=0$:
\begin{equation}\label{cc3}
    \tilde{n}_\mathrm{N}=\frac{e^{-\frac{E_{\mathrm{N}_\mathrm{O}^-}+ E_4}{k_B
    T}}}{\tilde{n}_4}.
\end{equation}
The energy $(E_{\mathrm{N}_\mathrm{O}^-}+ E_4)/2$ is given in Table \ref{t4}. In Fig.
\ref{f9} we present calculated  concentrations of Ti defects and Al vacancies at $T=T_m$
at the reference points of the $\mathrm{Al}-\mathrm{Ti}-\mathrm{O}-\mathrm{N}$ phase
diagram. One can see that the presence of nitrogen defects results in a considerable
increase of the concentration of $\mathrm{Ti}^{4+}$ and in an increase of the overall
concentration of Ti in the reduced and intermediate condition.

\begin{figure}
\includegraphics[width=0.4\linewidth]{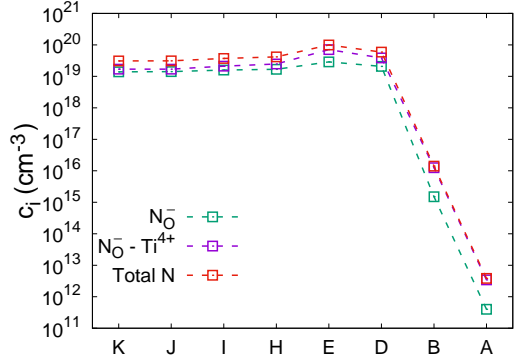}
\caption{ Equilibrium concentrations of nitrogen defects at $T=T_m$.} \label{f10}
\end{figure}

 Negatively charged nitrogen defects can bind in pairs
with $\mathrm{Ti}^{4+}$. The binding energy of such a pair is given in Table \ref{t3}.
The calculated concentration of isolated $\mathrm{N}_\mathrm{O}^-$ defects and
$\mathrm{N}_\mathrm{O}^--\mathrm{Ti}^{4+}$ pairs, and the total concentration of nitrogen
are shown in Fig. \ref{f10}. In the oxidized conditions the concentration of nitrogen
defects decreases and they do not influence the concentration of $\mathrm{Ti}^{4+}$. The
concentration of $\mathrm{N}_\mathrm{O}^--\mathrm{Ti}^{4+}$ pairs is slightly higher than
the concentration of isolated $\mathrm{N}_\mathrm{O}^-$ defects, but since these pairs
are electrically neutral, they  do not influence the concentration of charged defects.

The increase in the concentration of $\mathrm{Ti}^{3+}-\mathrm{Ti}^{4+}$ pairs results in
a decrease of the $c_{\mathrm{Ti}^{3+}}/c_{\mathrm{Ti}^{3+}-\mathrm{Ti}^{4+}}$ ratio and
a decrease of the FOM.

\subsection{$\mathrm{Al}-\mathrm{Ti}-\mathrm{O}-\mathrm{F}$ system}

Keeping in mind that fluorine ions enter into Ti:sapphire mostly in the form of
substitutional $\mathrm{F}_\mathrm{O}^+$ defects one would expect that co-doping with
fluorine decreases the concentration of $\mathrm{Ti}^{4+}$. But the situation appears to
be more complicated due to formation of negatively charged complexes of
$\mathrm{F}_\mathrm{O}^+$ and $V_{\mathrm{Al}}^{3-}$.

\begin{figure}
\includegraphics[width=0.5\linewidth]{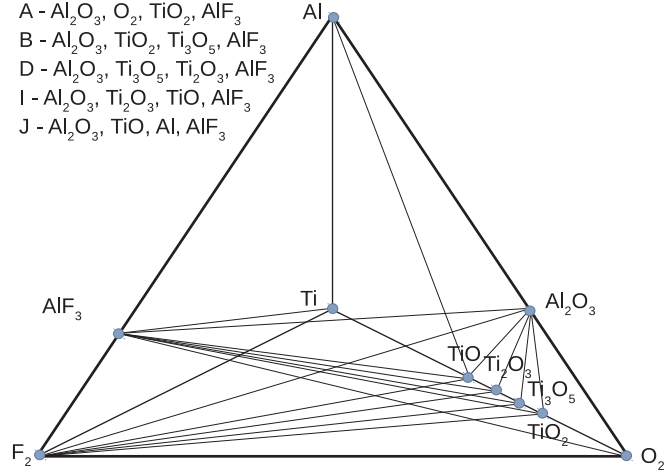}
\caption{The phase diagram of the Al$-$Ti$-$O$-$F system at $T=T_m$ and reference
points.} \label{f11}
\end{figure}

The calculated four-component $\mathrm{Al}-\mathrm{Ti}-\mathrm{O}-\mathrm{F}$
 phase diagram at $T=T_m$ is shown in Fig. \ref{f11}.
 This diagram contains five reference points (A, B, D, I, J).
 In all these points Al$_2$O$_3$ is in equilibrium with AlF$_3$. Two other compounds are the same
  as for the reference points of the  $\mathrm{Al}-\mathrm{Ti}-\mathrm{O}$ diagram (Fig. \ref{f1}).

Calculations show that together with positively charged $\mathrm{F}_\mathrm{O}^+$ defects
two types of negatively charged complexes formed by fluorine have small formation
energies. They are the pairs $\mathrm{F}_\mathrm{O}^+-V_\mathrm{Al}^{3-}$ and the triples
$\mathrm{F}_\mathrm{O}^+-\mathrm{F}_\mathrm{O}^+-V_\mathrm{Al}^{3-}$.  To calculate
equilibrium concentrations of defects we take into account three fluorine defect species
and add to the left-hand-part of Eq. (\ref{cn2}) the terms $ ({3}/{2}) \tilde{n}_{F}- 12
C_{\mathrm{F}V}\tilde{n}_\mathrm{F}
 \tilde{n}_V- 15 C_{\mathrm{F}V\mathrm{F}}\tilde{n}_\mathrm{F}^2
\tilde{n}_V$, where $ \tilde{n}_\mathrm{F}$ is the number of $\mathrm{F}_\mathrm{O}^+$
defects normalized to the total number of oxygen atoms. The quantities
$\tilde{n}_\mathrm{F}$ and $\tilde{n}_4$ satisfy the equation
\begin{equation}\label{201} \tilde{n}_\mathrm{F}=\tilde{n}_4
e^{-\frac{E_{\mathrm{F}_\mathrm{O}^+}-E_4}{k_B T}},
\end{equation}
which follows from Eq. (\ref{7}). The difference $E_{\mathrm{F}_\mathrm{O}^+}-E_4$ is
given in Table \ref{t4}. The binding energies of the complexes
$\mathrm{F}_\mathrm{O}^+-V_\mathrm{Al}^{3-}$ and
$\mathrm{F}_\mathrm{O}^+-\mathrm{F}_\mathrm{O}^+-V_\mathrm{Al}^{3-}$ are presented in
Table \ref{t3}.

\begin{figure}
\includegraphics[width=0.4\linewidth]{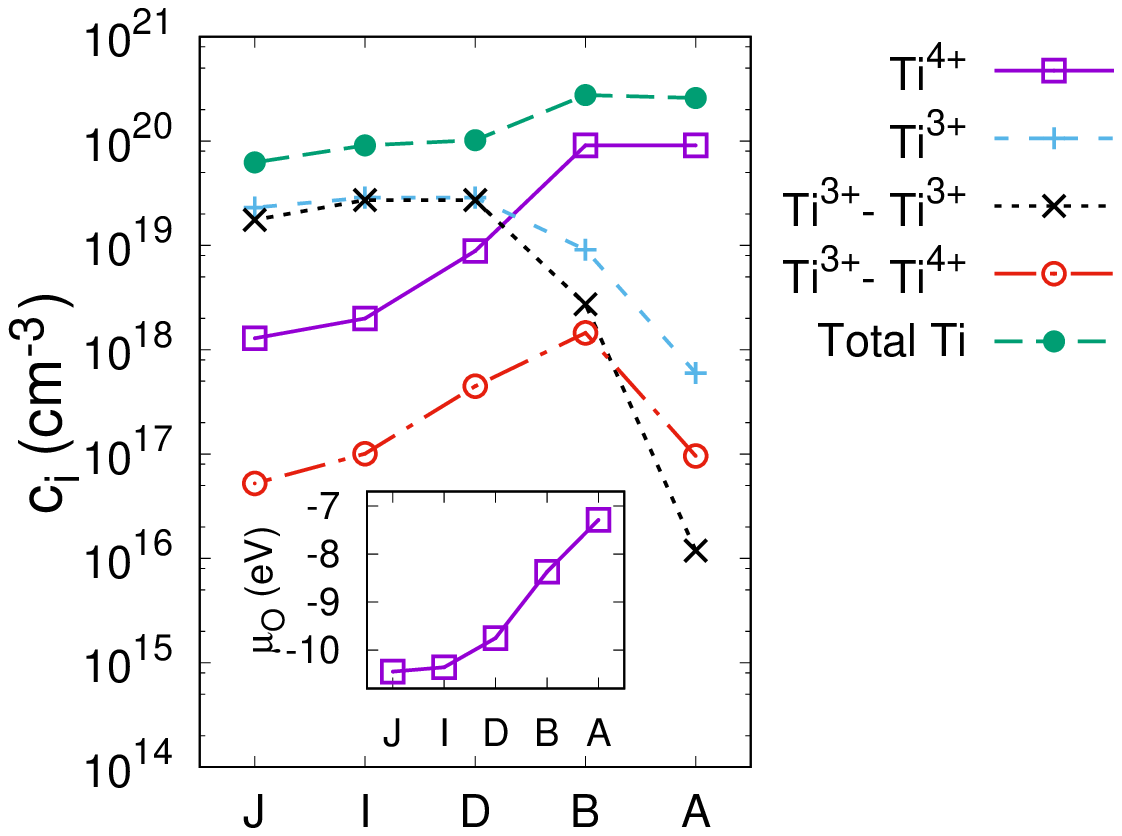}
\includegraphics[width=0.47\linewidth]{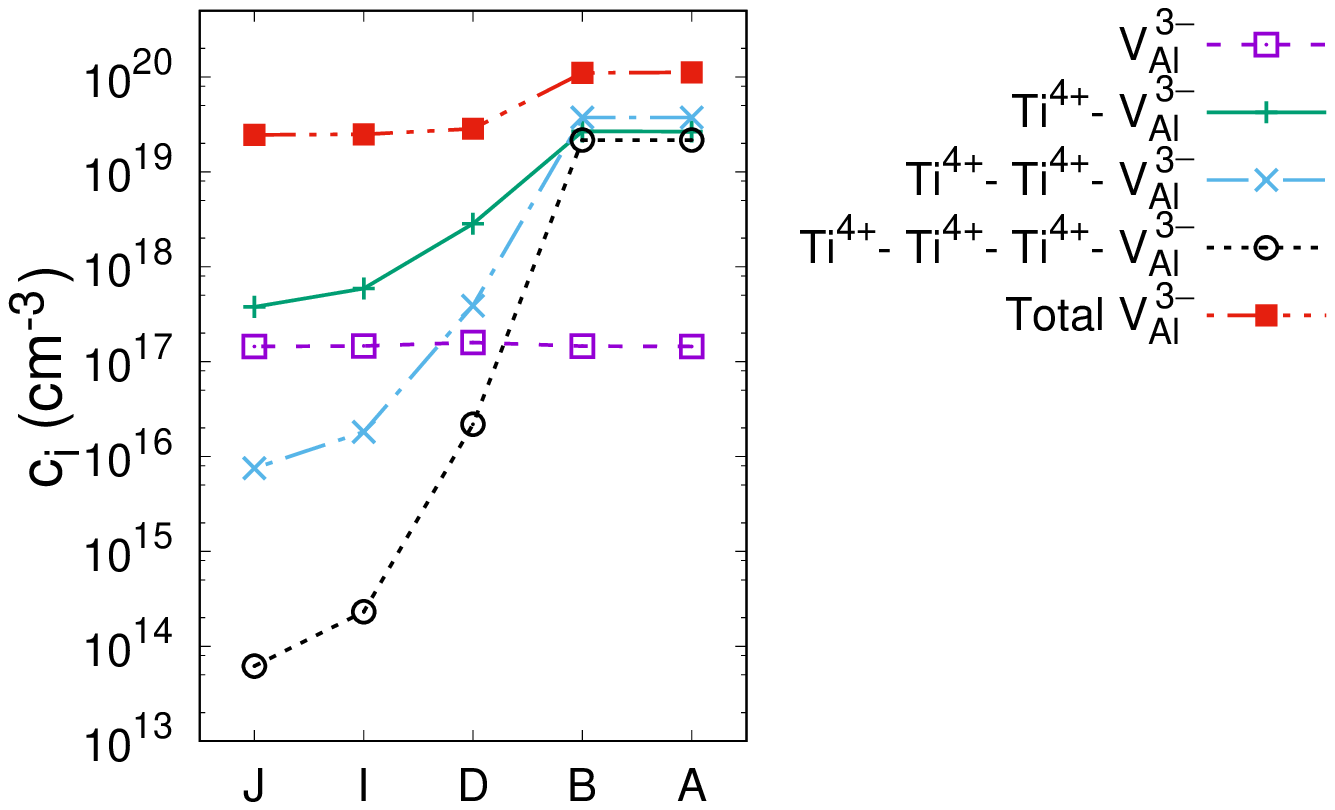}
\caption{ Equilibrium concentrations of defects in Ti:sapphire at $T=T_m$  in the
conditions that correspond to reference points of
$\mathrm{Al}-\mathrm{Ti}-\mathrm{O}-\mathrm{F}$ phase diagram Fig. \ref{f11}.}
\label{f12}
\end{figure}

\begin{figure}
\includegraphics[width=0.4\linewidth]{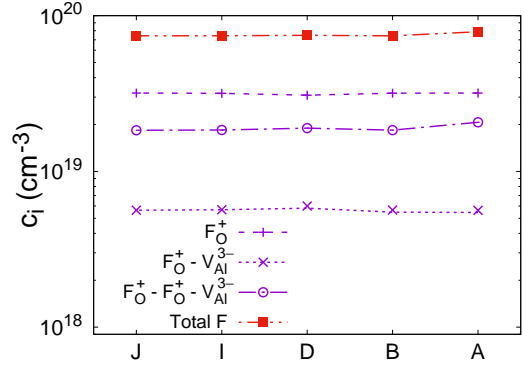}
\caption{ Equilibrium concentrations of fluorine defects at $T=T_m$.} \label{f13}
\end{figure}

The calculated equilibrium concentrations of defects are shown in Figs. \ref{f12} and
\ref{f13}. One can see from Fig. \ref{f13} that the concentrations of positively charged
$\mathrm{F}_\mathrm{O}^+$ and  negatively charged
$\mathrm{F}_\mathrm{O}^+-\mathrm{F}_\mathrm{O}^+-V_\mathrm{Al}^{3-}$ are close to each
other. Negatively charged defects almost compensate positively charged ones and the
concentrations of $\mathrm{Ti}^{4+}$ and $\mathrm{Ti}^{3+}-\mathrm{Ti}^{4+}$ are changed
unessentially. Thus, our expectation on a positive role of fluorine is not confirmed. It
appears that triples $\mathrm{F}_\mathrm{O}^+-\mathrm{F}_\mathrm{O}^+-V_\mathrm{Al}^{3-}$
have rather small formation energy and the positive effect caused by isolated
$\mathrm{F}_\mathrm{O}^+$ ions reduces due to formation of
$\mathrm{F}_\mathrm{O}^+-\mathrm{F}_\mathrm{O}^+-V_\mathrm{Al}^{3-}$ triples. We note
that such a reduction is not a general rule. It would not happen if the binding energy of
complexes was smaller.

\subsection{Influence of co-doping on FOM} \label{fom}

\begin{figure}
\includegraphics[width=0.5\linewidth]{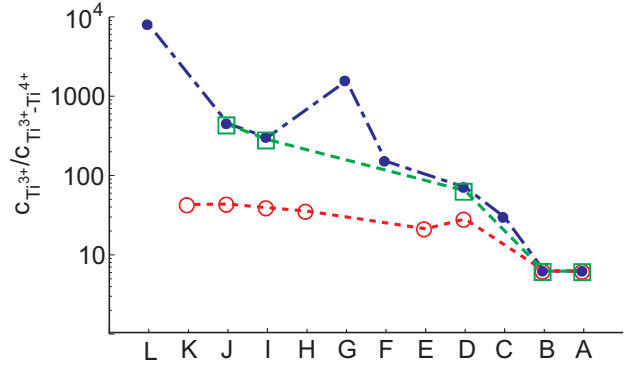}
\caption{The ratio of concentrations of Ti$^{3+}$ to the concentration of
$\mathrm{Ti}^{3+}-\mathrm{Ti}^{4+}$ pairs at reference points of Al$-$Ti$-$O$-X$ phase
diagrams (see Table \ref{t1}) at $T=T_m$. Filled circles correspond to $X=\mathrm{C}$,
open circles, to $X=\mathrm{N}$ and open squares, to  $X=\mathrm{F}$. Lines are guides to
the eye.} \label{f14}
\end{figure}

To determine the impact of carbon, nitrogen and fluorine compounds on the FOM of
Ti:sapphire we calculate the $c_{\mathrm{Ti}^{3+}}/c_{\mathrm{Ti}^{3+}-\mathrm{Ti}^{4+}}$
ratio at all reference points of Table \ref{t1} using the data presented in Figs.
\ref{f6}, \ref{f9} and \ref{f12}. The result is shown in Fig. \ref{f14}. Since the
concentration of carbon defects is low, at the points A, B, D, I ,J the result is the
same for Al$-$Ti$-$O$-$C and Al$-$Ti$-$O systems. At the same time under conditions that
corresponds to the points G and L one can to reach the much larger ratio of
$c_{\mathrm{Ti}^{3+}}/c_{\mathrm{Ti}^{3+}-\mathrm{Ti}^{4+}}$. The latter is connected
with the low total equilibrium concentration of Ti at these points. This conclusion is in
agreement with experimental study \cite{nizh1} where it was shown that thermal carbon
treatment of raw materials (Al$_2$O$_3$ and TiO$_2$) makes it possible to decrease the
concentration of Ti$^{4+}$ in Ti:Al$_2$O$_3$ crystals. With this, Fig. \ref{f14}
illustrates that nitrogenization provokes considerable reduction of the
$c_{\mathrm{Ti}^{3+}}/c_{\mathrm{Ti}^{3+}-\mathrm{Ti}^{4+}}$ ratio in the reduced and
intermediate conditions, and that fluoridization leaves this ratio almost unchanged.

It is instructive to evaluate the relation between the FOM and
$c_{\mathrm{Ti}^{3+}}/c_{\mathrm{Ti}^{3+}-\mathrm{Ti}^{4+}}$. To do that we use the data
of \cite{m1}. One of the samples investigated in \cite{m1} (labeled as SY1b) had
$\mathrm{FOM}=12$. Concentrations of $\mathrm{Ti}^{3+}$ and $\mathrm{Ti}^{4+}$ in this
sample were estimated as $1.3\cdot 10^{18}$ cm$^{-3}$ and $2.3\cdot 10^{18}$ cm$^{-3}$,
correspondingly. Using Eq. (\ref{13}) and taking $T=2000$ K we obtain
$c_{\mathrm{Ti}^{3+}-\mathrm{Ti}^{4+}}\approx 10^{16}$ cm$^{-3}$ and
$c_{\mathrm{Ti}^{3+}}/c_{\mathrm{Ti}^{3+}-\mathrm{Ti}^{4+}}\approx 100$. Thus one can
estimate that $\mathrm{FOM}\sim 0.1\cdot
c_{\mathrm{Ti}^{3+}}/c_{\mathrm{Ti}^{3+}-\mathrm{Ti}^{4+}}$. Note that, normally,
commercial samples have FOM $=100 \div 300$ and FOM of the best samples is up to 1000.

\subsection{Band structure of Ti:sapphire co-doped with carbon, nitrogen and fluorine}

 Co-doping may result in the appearance of additional defect levels in the band gap. To
 consider this effect we calculate the band structure of the system with one given defect per
 supercell. Strictly speaking, from such calculations one obtains impurity bands of a crystal
 with periodically
 arranged defects.
 But since such bands are very narrow they can be associated with impurity levels
 connected with a given defect. The position of obtained impurity levels in the band gap and
 separation between them  weakly depend on the supercell size.
 This can be seen from comparison of the results
 of Refs. \cite{lv1,lv2} and our calculations \cite{my}.

\begin{figure}
\begin{minipage}{0.3\linewidth}
\center{\includegraphics[width=1\linewidth]{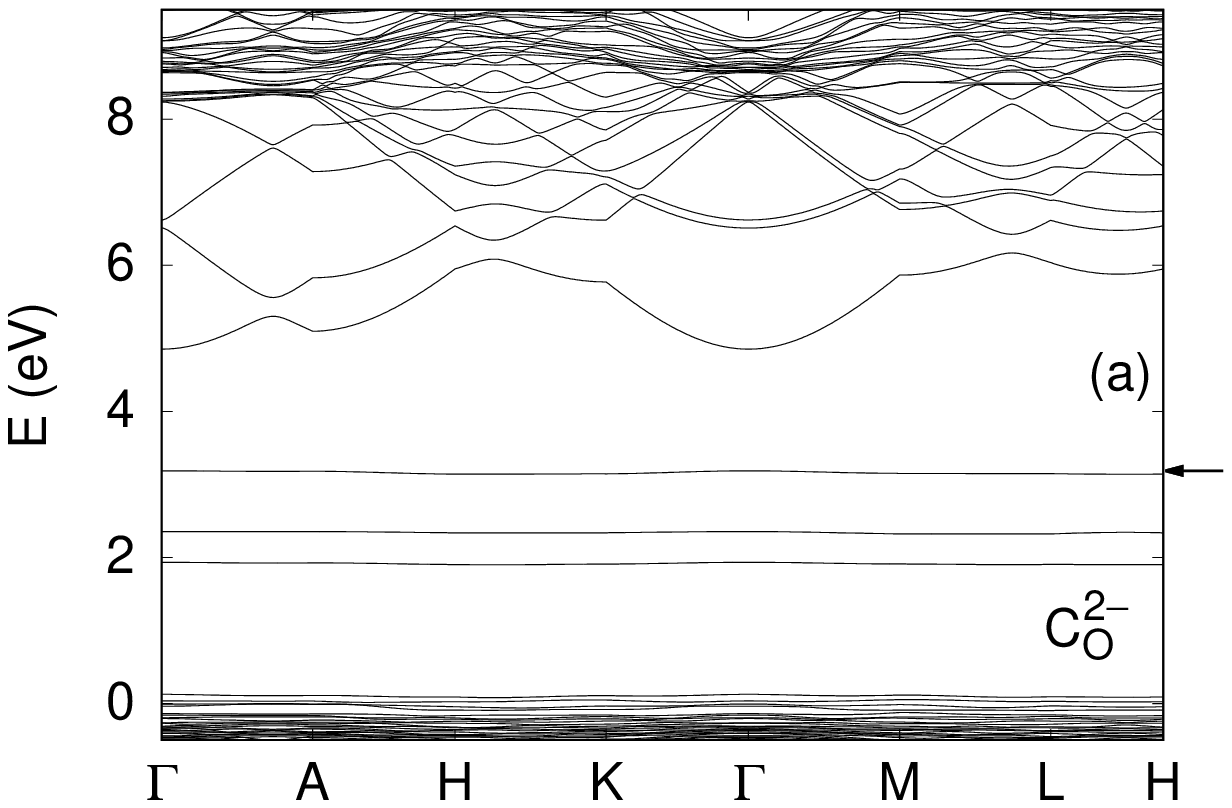}}
\end{minipage}\\
\begin{minipage}{0.3\linewidth}
\center{\includegraphics[width=1\linewidth]{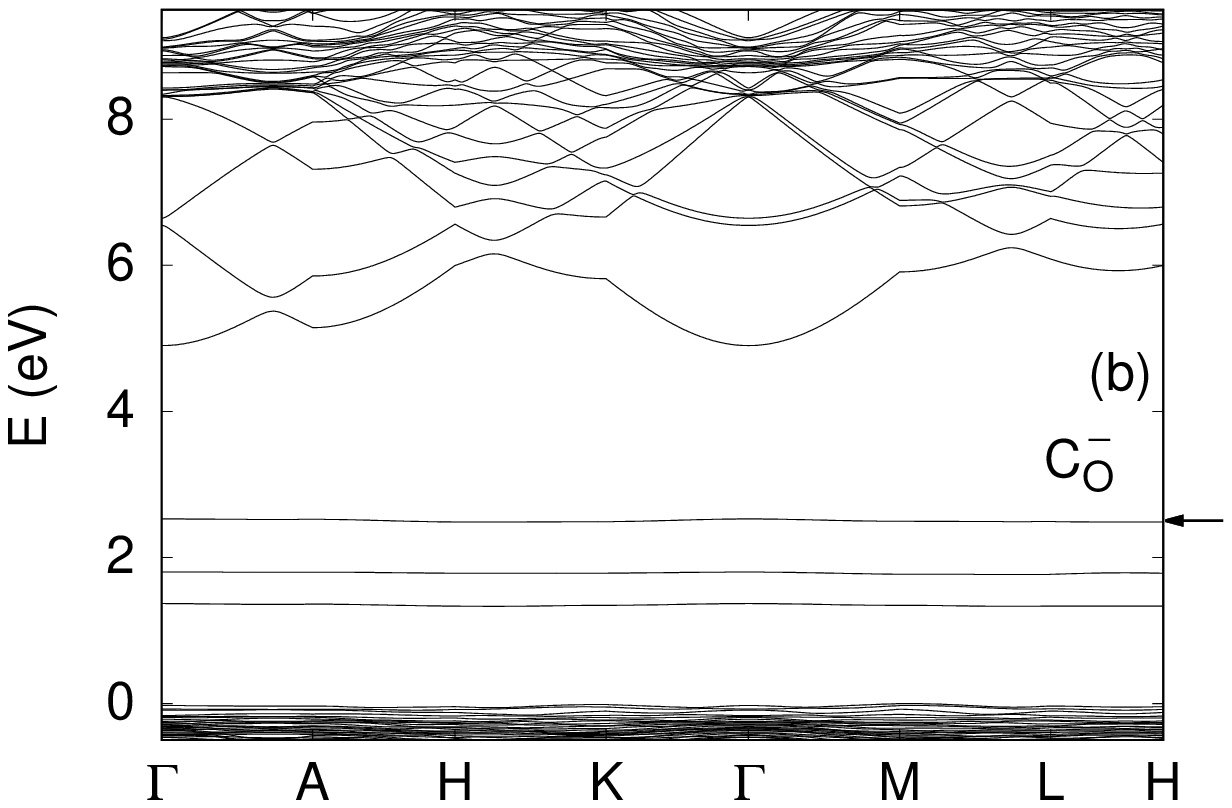}}
\end{minipage}\\
\begin{minipage}{0.3\linewidth}
\center{\includegraphics[width=1\linewidth]{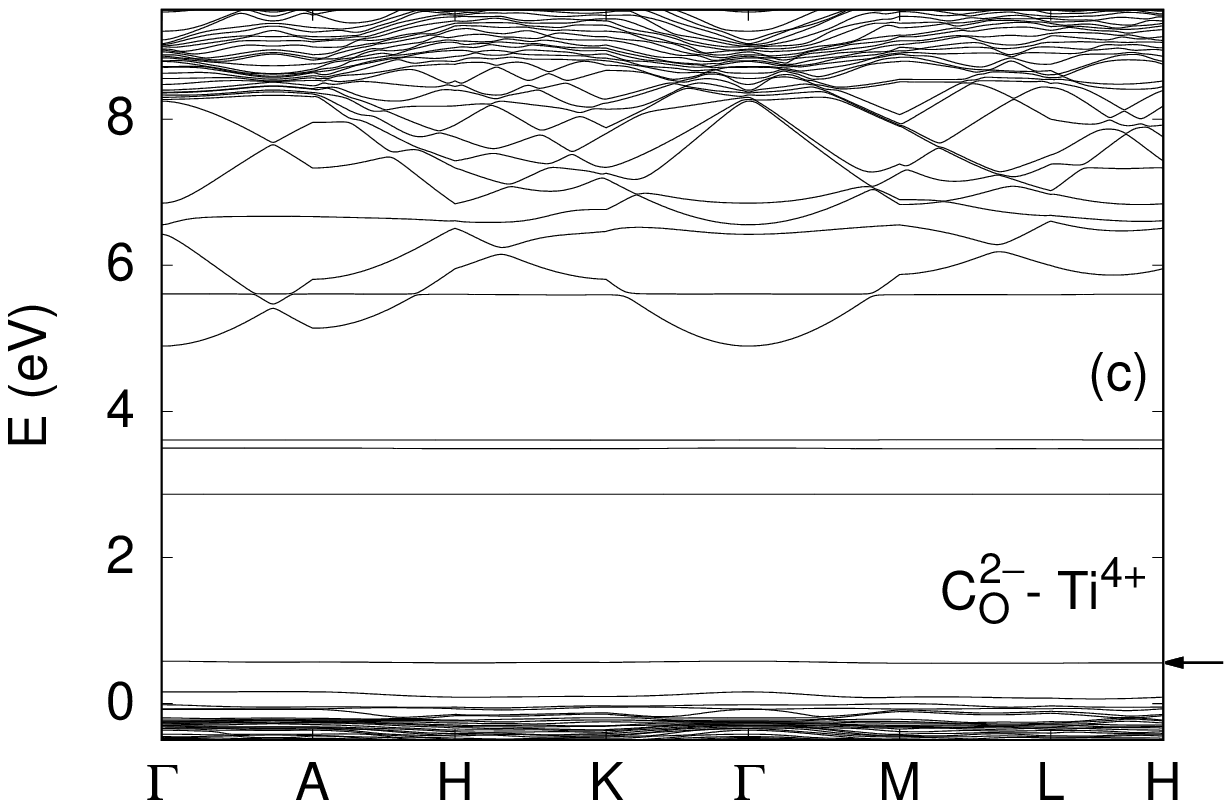}}
\end{minipage}\\
\begin{minipage}{0.3\linewidth}
\center{\includegraphics[width=1\linewidth]{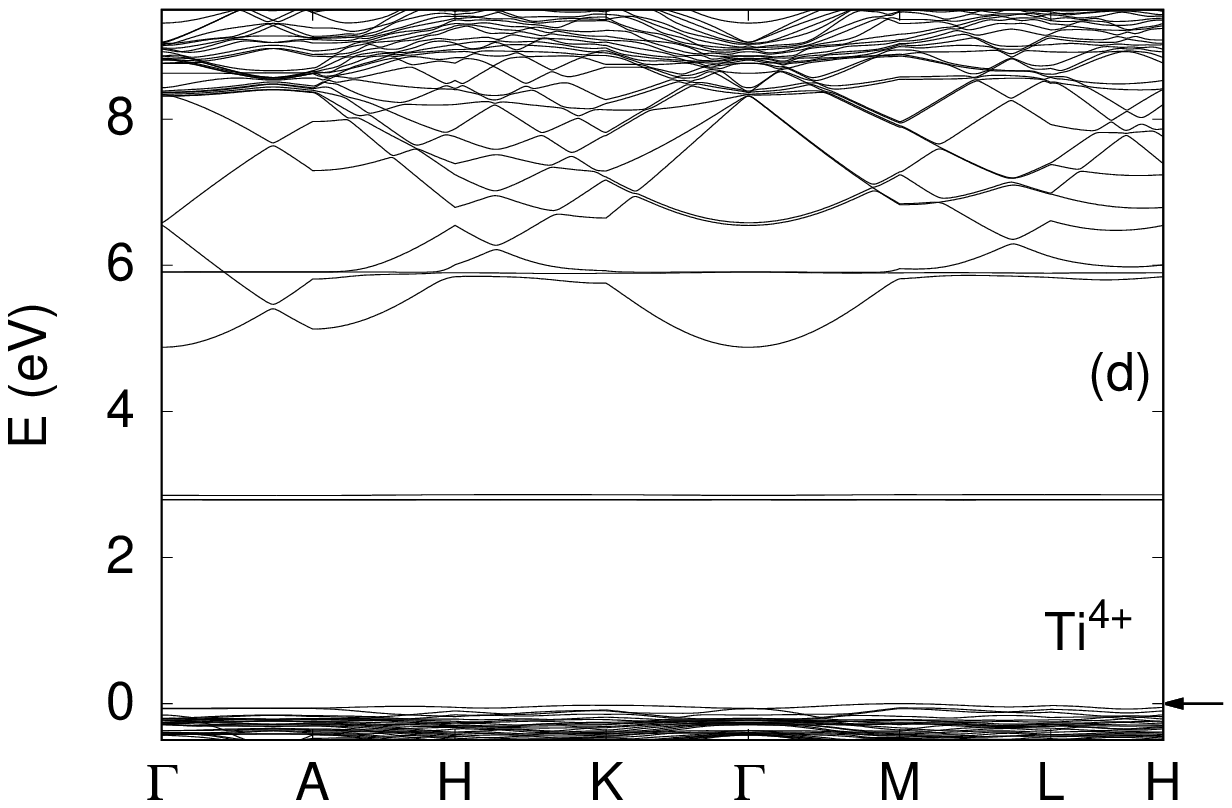}}
\end{minipage}
\caption{Band structure of $\mathrm{Al}_2\mathrm{O}_3$ with $\mathrm{C}_\mathrm{O}^{2-}$
(a), $\mathrm{C}_\mathrm{O}^{-}$  (b), $\mathrm{C}_\mathrm{O}^{2-}-\mathrm{Ti}^{4+}$ (c)
and $\mathrm{Ti}^{4+}$ (d) defects. The valence band maximum is set at 0 eV, and the
arrow indicates the position of the highest occupied level.} \label{f15}
\end{figure}

\begin{figure}
\begin{minipage}{0.3\linewidth}
\center{\includegraphics[width=1\linewidth]{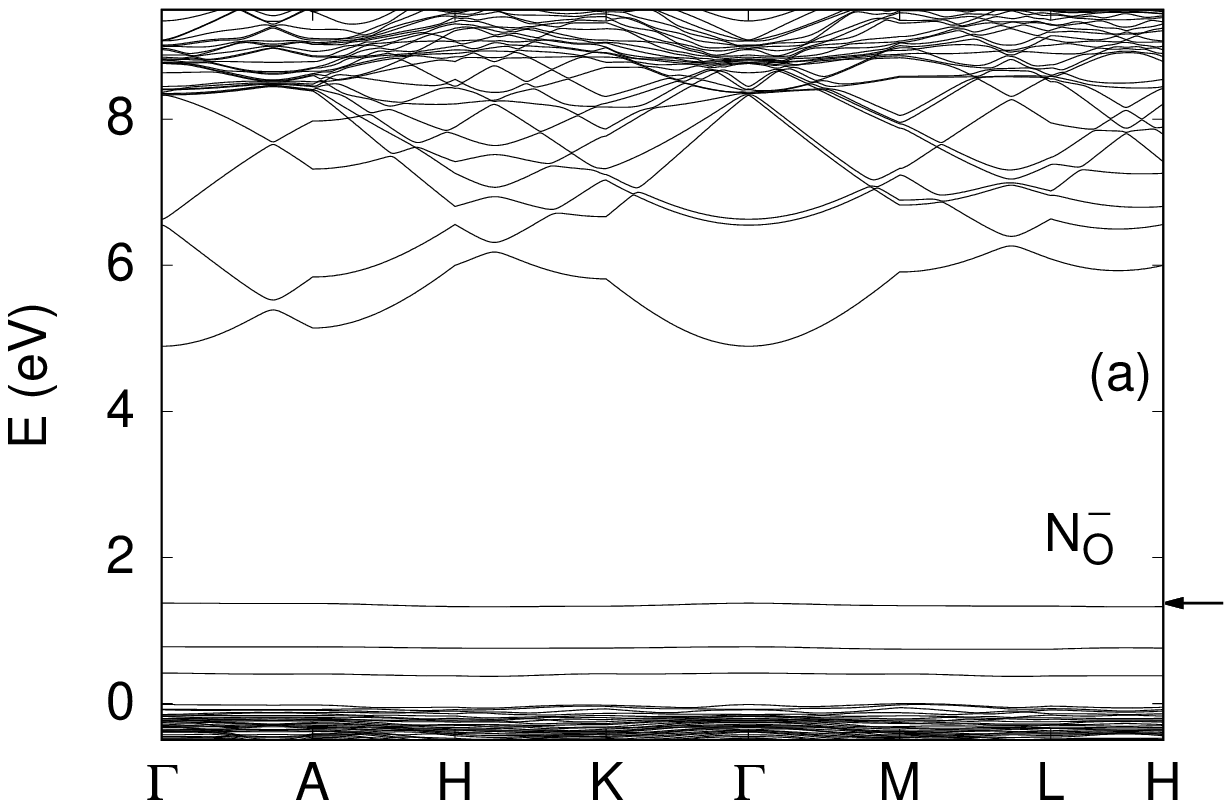}}
\end{minipage}\\
\begin{minipage}{0.3\linewidth}
\center{\includegraphics[width=1\linewidth]{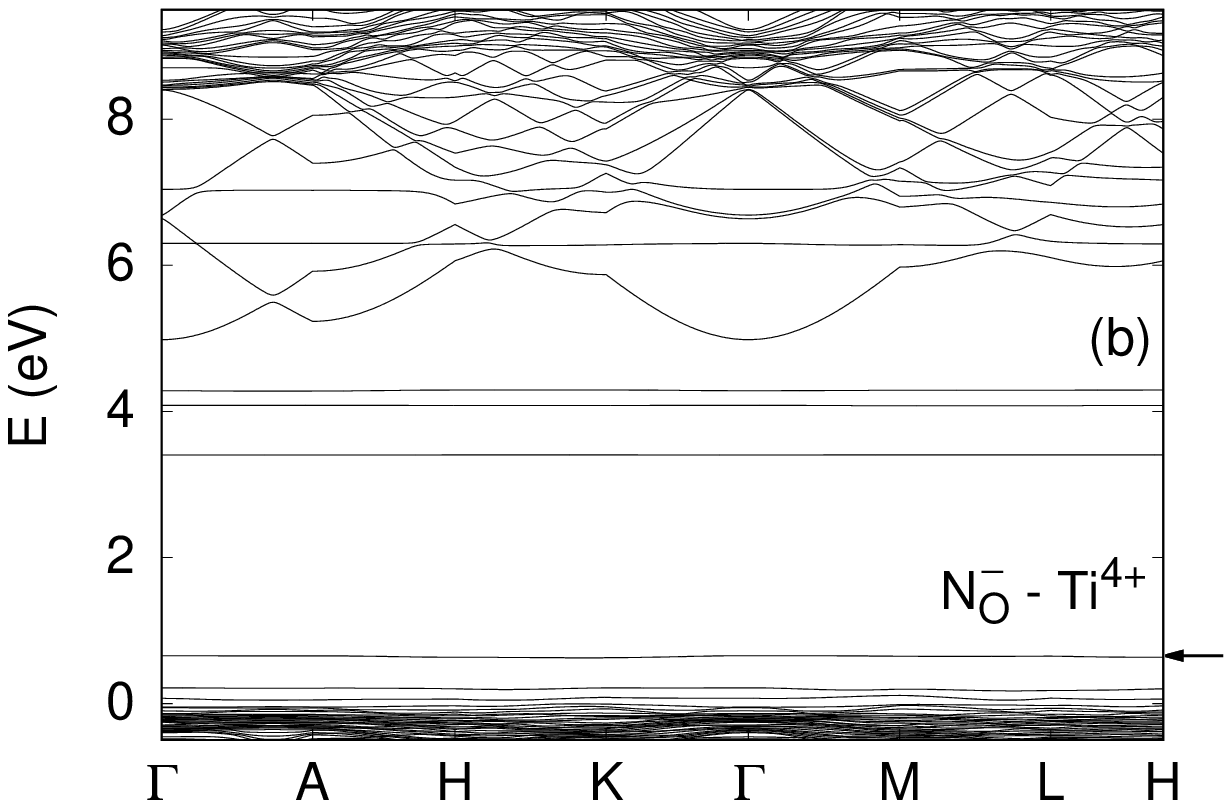}}
\end{minipage}
\caption{The same as in Fig. \ref{f15} for  $\mathrm{N}_\mathrm{O}^-$  (a) and
$\mathrm{N}_\mathrm{O}^{-}-\mathrm{Ti}^{4+}$ (b) } \label{f16}
\end{figure}

\begin{figure}
\begin{minipage}{0.3\linewidth}
\center{\includegraphics[width=1\linewidth]{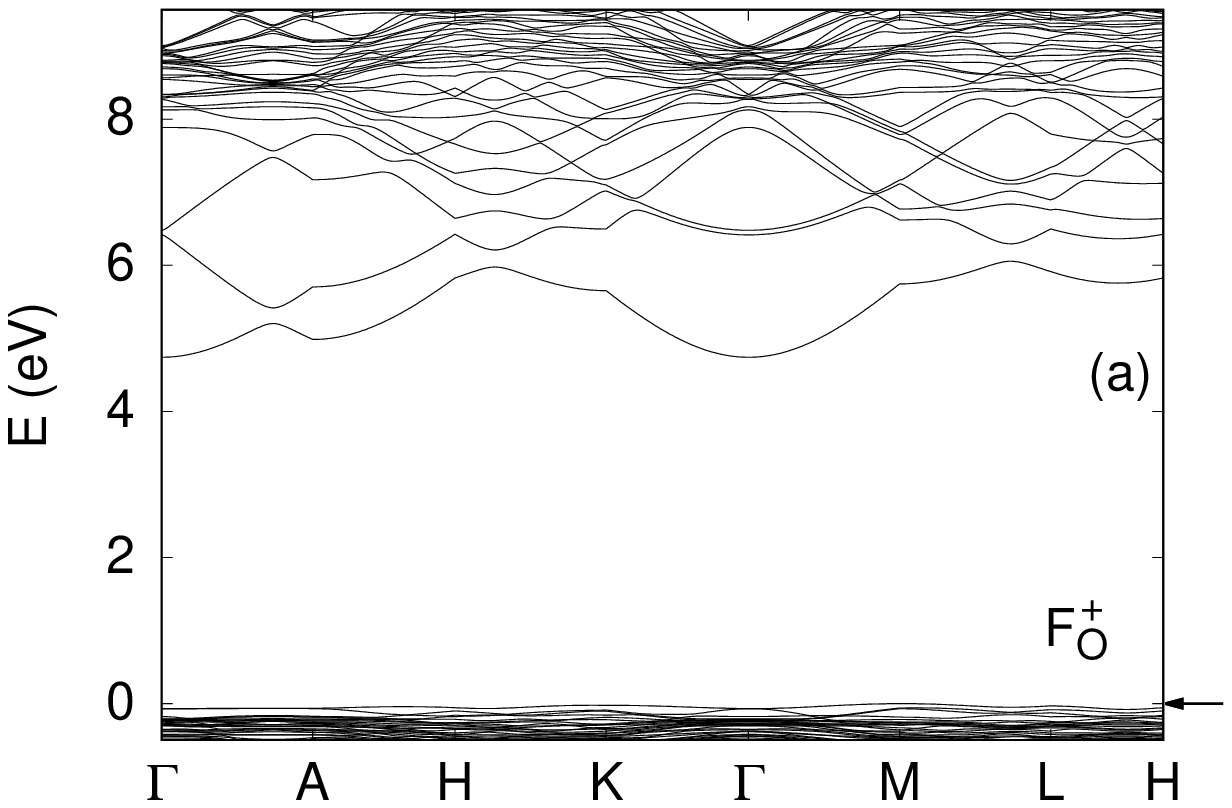}}
\end{minipage}\\
\begin{minipage}{0.3\linewidth}
\center{\includegraphics[width=1\linewidth]{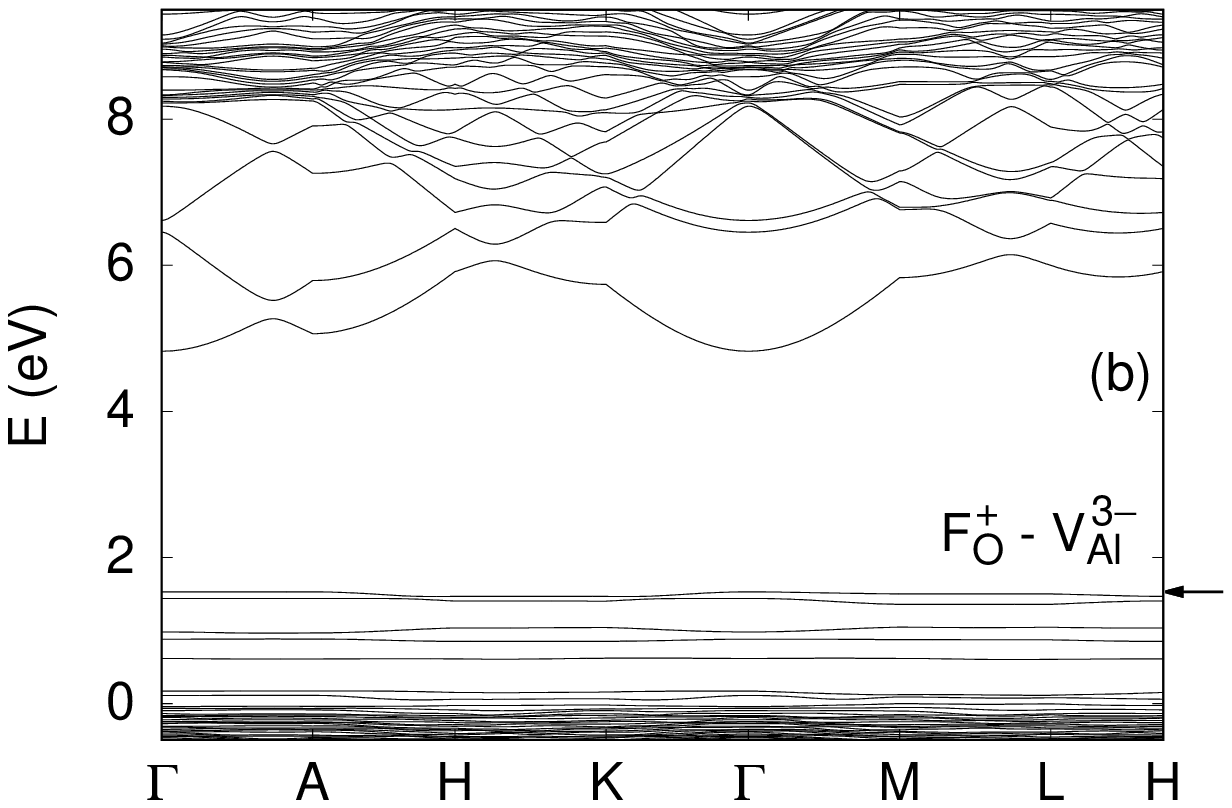}}
\end{minipage}\\
\begin{minipage}{0.3\linewidth}
\center{\includegraphics[width=1\linewidth]{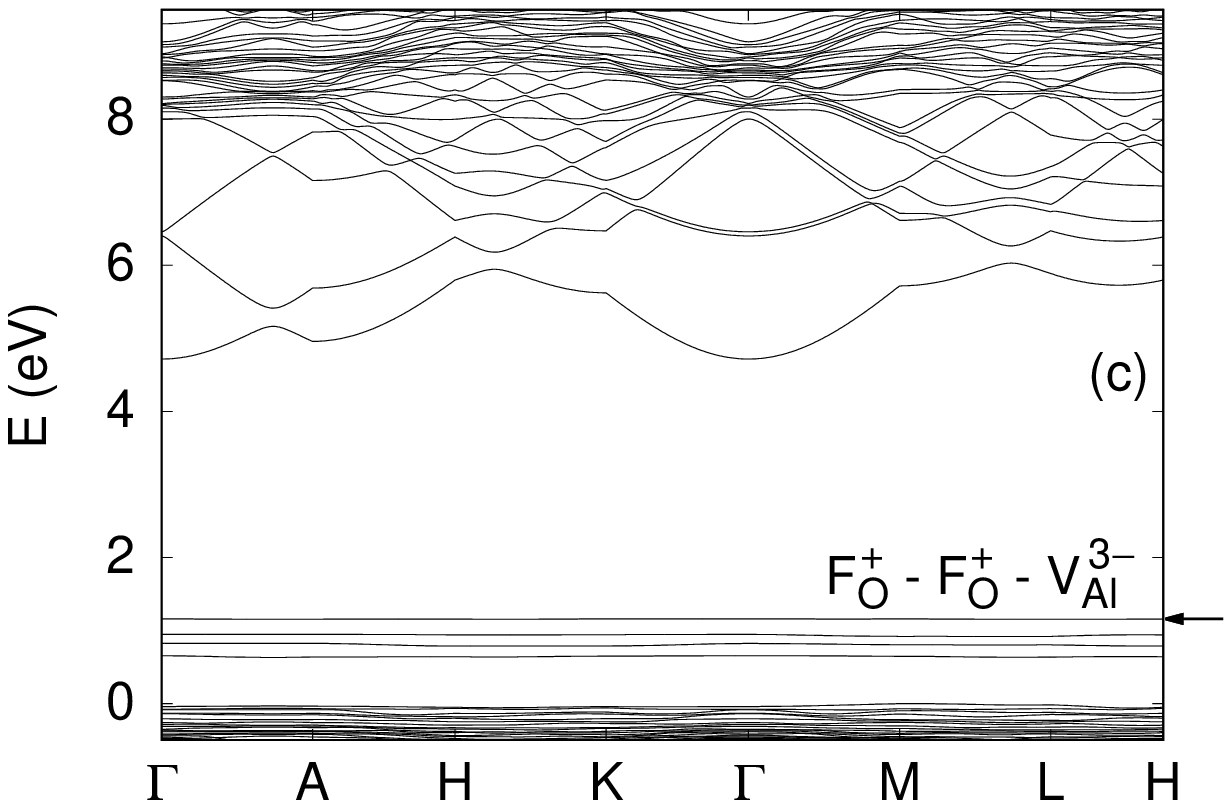}}
\end{minipage}\\
\begin{minipage}{0.3\linewidth}
 \center{\includegraphics[width=1\linewidth]{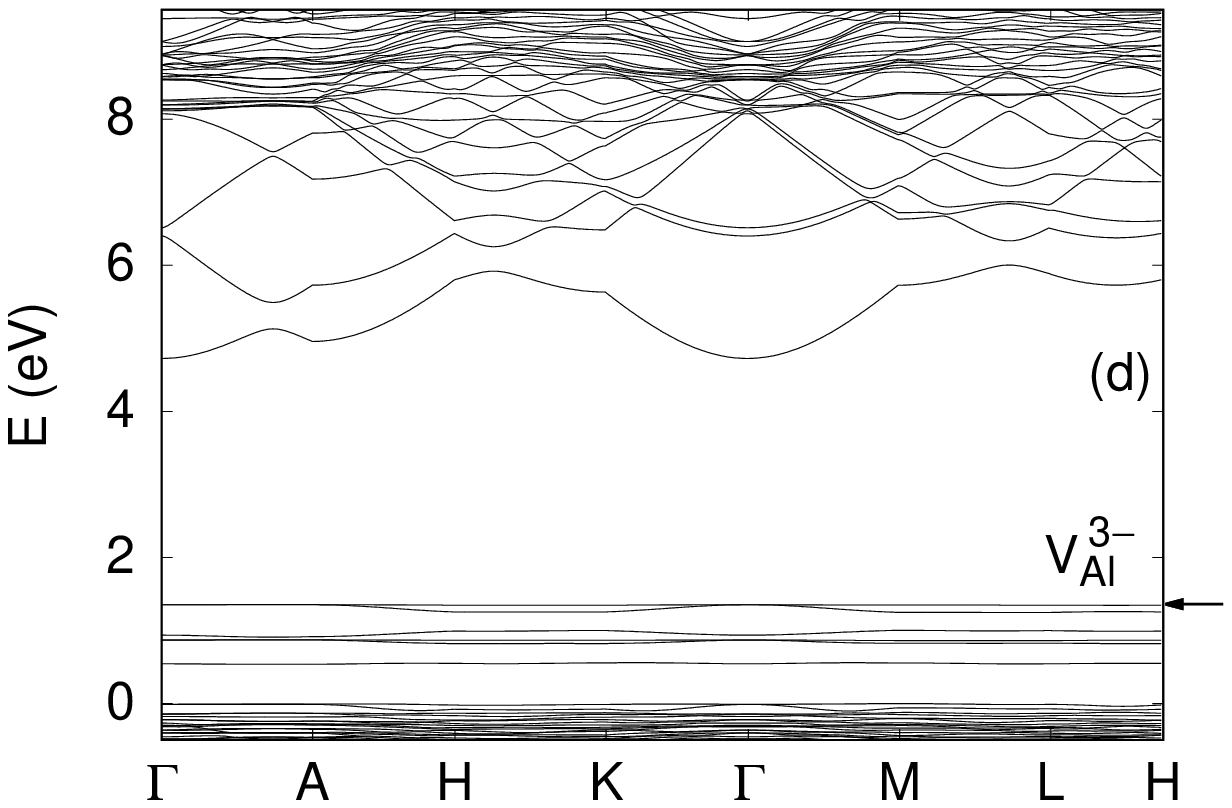}}
\end{minipage}
\caption{The same as in Fig. \ref{f15} for a $\mathrm{F}_\mathrm{O}^+$ (a),
$\mathrm{F}_\mathrm{O}^{+}-V_\mathrm{Al}^{3-}$ (b),
$\mathrm{F}_\mathrm{O}^{+}-\mathrm{F}_\mathrm{O}^{+} -V_\mathrm{Al}^{3-}$ (c) and
$V_\mathrm{Al}^{3-}$ (d)} \label{f17}
\end{figure}

 In Figs. \ref{f15}, \ref{f16} and \ref{f17} we present the band structure of a sapphire
  crystal with carbon, nitrogen and fluorine defects. For comparison, in
  Fig. \ref{f15}d and Fig. \ref{f17}d we reproduce the
band structures of  crystals   with $\mathrm{Ti}^{4+}$ substitutional defects and with
  $V_\mathrm{Al}^{3-}$ vacancies \cite{my}.
One can see that fluorine defects do not cause additional impurity levels in the band
gap. Formation of complexes of fluorine defects with Al vacancy results in a minor
modification of Al vacancy levels. In contrast, negatively charged carbon and nitrogen
defects ($\mathrm{C}_\mathrm{O}^{2-}$,
  $\mathrm{C}_\mathrm{O}^{-}$ and $\mathrm{N}_\mathrm{O}^{-})$ reveal themselves in
  an appearance of additional impurity levels. Binding of such defects with
  $\mathrm{Ti}^{4+}$ causes splitting of Ti impurity levels.

Due to small equilibrium
  concentration of carbon defects one can expect that carbon impurities will not
  influence significantly optical properties of Ti:sapphire.
   Equilibrium   concentration of nitrogen defects is much higher.  Therefore
  growth or annealing in the presence of nitrogen compounds may result in an essential
  modification of optical properties of Ti:sapphire not only due to a change of the balance between
  $\mathrm{Ti}^{3+}$ and $\mathrm{Ti}^{4+}$ ions, but also due to the appearance of
  impurity levels caused by  nitrogen defects.

\section{Universal relations between the concentrations of $\mathrm{Ti}^{3+}$, $\mathrm{Ti}^{4+}$
and $\mathrm{Ti}^{3+}-\mathrm{Ti}^{4+}$} \label{s6}

In this section we discuss  whether the observed relations between the concentrations of
isolated $\mathrm{Ti}^{3+}$ and $\mathrm{Ti}^{4+}$ defects, and
$\mathrm{Ti}^{3+}-\mathrm{Ti}^{4+}$ pairs confirm or put in question the pair model of
NIR absorption.

\begin{figure}
\includegraphics[width=0.4\linewidth]{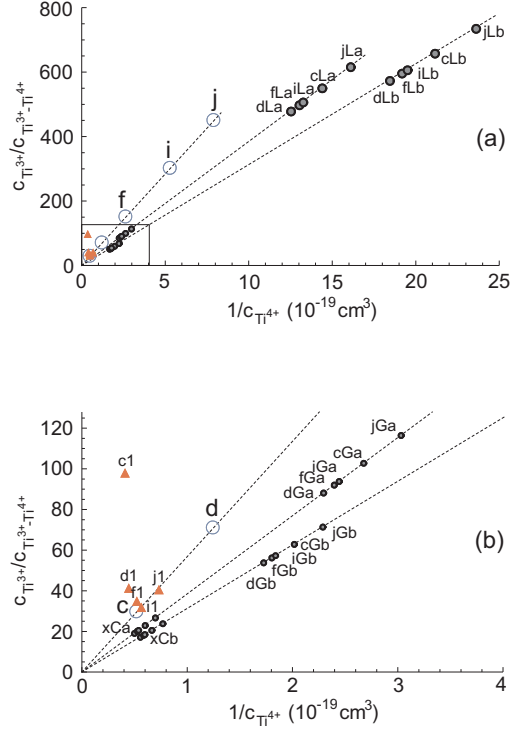}
\caption{The  $c_{\mathrm{Ti}^{3+}}/c_{\mathrm{Ti}^{3+}-\mathrm{Ti}^{4+}}$ ratio versus
the inverse concentration of $\mathrm{Ti}^{4+}$. Open circles show equilibrium
concentrations at $T=T_m$,  filled circles, at $T=2100$K and $T=2000$K,  labeled as xYa
and xYb, respectively ($\mathrm{x}=\mathrm{c,d,f,i,j}$ refers to the parent sample, and
$\mathrm{Y}=\mathrm{C,G,L}$, to the reference point in the Al$-$O$-$C phase diagram).
Triangles correspond to equilibrium concentration at $T=1600$ K reached under condition
that diffusion of Ti is suppressed. Dashed lines show linear fit. A rectangle in (a)
indicates the area shown in (b) in another scale.} \label{f18}
\end{figure}

According to Eq. (\ref{13}), for samples where the equilibrium concentration of defects
was reached at the same temperature,
 e.g. the samples are annealed at the same $T$,  the factor
$\exp(-E^{(b)}_{3-4}/k_B T)$ will be the same and the ratio of the concentration of
$\mathrm{Ti}^{3+}$ to the concentration of $\mathrm{Ti}^{3+}-\mathrm{Ti}^{4+}$ will be in
inverse proportion to the concentration of $\mathrm{Ti}^{4+}$.
 Assuming that the
$c_{\mathrm{Ti}^{3+}}/c_{\mathrm{Ti}^{3+}-\mathrm{Ti}^{4+}}$ ratio determines the FOM one
can expect that the FOM would be in inverse proportion to $c_{\mathrm{Ti}^{4+}}$ as well.
At the same time samples annealed at different temperatures will have different
coefficients of proportionality between these quantities and their FOM may not
demonstrate such a proportionality. Let us consider the following example. We  imagine
that someone obtained five samples labeled as c, d, f, i and j. These samples were grown
in the conditions that correspond to the reference points C, D, F, I and J of the
Al$-$Ti$-$O$-$C phase diagram, respectively. According to our calculations (see Fig.
\ref{f6}) these samples should have approximately the same total concentration of Ti, but
different concentrations of $\mathrm{Ti}^{4+}$. Then we imagine that each sample was
divided into seven parts and 30 samples were annealed  in the conditions that correspond
to the reference points C, G and L of the $\mathrm{Al}-\mathrm{Ti}-\mathrm{C}$ phase
diagram at two different temperatures, $T=2100$ K and $T=2000$ K.  The reference points
for the $\mathrm{Al}-\mathrm{Ti}-\mathrm{C}$ diagram and the value of $\mu_\mathrm{O}$ at
three different temperatures are given in Table \ref{t5} (under lowering in temperature
two reference points, C and G, disappear and are replaced with the point C$_1$). We imply
that the total concentration of Ti is not changed under annealing. As a result there were
obtained 35 samples: 5 parent samples c, d, f, i and j; 15 samples annealed at $T=2100$ K
in the conditions that correspond to the points C, G and L of the Al$-$O$-$C phase
diagram  and  labeled as xCa, xGa, and xLa (x=c,d,f,i,j); and 15 samples annealed at
$T=2000$ K and labeled as xCb, xGb, and xLb, correspondingly). We calculate the
$c_{\mathrm{Ti}^{3+}}/c_{\mathrm{Ti}^{3+}-\mathrm{Ti}^{4+}}$ ratio for these samples and
plot it against $1/c_{\mathrm{Ti}^{4+}}$ (Fig. \ref{f18}). One can see that the points in
Fig. \ref{f18} belong to three zero origin straight lines with different slopes. If one
takes randomly several samples from this 35-sample set and measures the FOM one may find
that the FOM is not proportional to $1/c_{\mathrm{Ti}^{4+}}$. But this does not mean that
the pair model of NIR absorption is incorrect. This example demonstrates that it is
important to investigate samples annealed at the same temperature  to verify the pair
model of NIR absorption.

\begin{table}\centering
  \caption{Oxygen chemical potential at reference points of $\mathrm{Al}-\mathrm{O}-\mathrm{C}$
  phase diagram at three different temperatures.}
  \label{t5}
\begin{tabular}{|c|c|c|c|c|c|}
\hline \multirow{2}{*}{$T$,K} & \multicolumn{5}{c|}{Reference points, phases in
equilibrium, $\mu_\mathrm{O}$ in eV}
\\ \cline{2-6}
                   & \begin{tabular}[c]{@{}c@{}}A\\ $\mathrm{Al}_2\mathrm{O}_3$, $\mathrm{O}_2$, $\mathrm{CO}_2$
                   \end{tabular} & \begin{tabular}[c]{@{}c@{}}C\\
                   $\mathrm{Al}_2\mathrm{O}_3$,  $\mathrm{CO}_2$, CO
                   \end{tabular}  &  \begin{tabular}[c]{@{}c@{}}C$_1$\\
                   $\mathrm{Al}_2\mathrm{O}_3$,  $\mathrm{CO}_2$, C
                   \end{tabular}  & \begin{tabular}[c]{@{}c@{}}G\\
                   $\mathrm{Al}_2\mathrm{O}_3$,  CO, C
                   \end{tabular}   & \begin{tabular}[c]{@{}c@{}}L\\
                   $\mathrm{Al}_2\mathrm{O}_3$,  C, Al
                   \end{tabular}   \\ \hline
2100               & -6.97 & -9.11  & -  & -9.73  & -10.38 \\
\hline
2000               &-6.83 & -9.06  &-  & -9.50 & -10.35 \\
\hline 1600          &-6.28 & - &-8.73  & -  & -10.25 \\
 \hline
\end{tabular}
\end{table}

The dependencies presented in Fig. \ref{f18} also demonstrate that annealing in the
reduced conditions (point L) should increase the FOM and  the effect is larger at lower
temperature of annealing. At the same time annealing in the intermediate conditions
(points G and C) may decrease the FOM.

\begin{figure}
\includegraphics[width=0.4\linewidth]{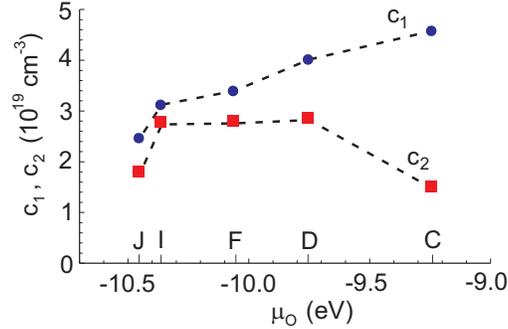}
\caption{Sum of equilibrium  concentrations of isolated Ti and pairs
Ti$^{4+}-V_\mathrm{Al}$ ($c_1$) and sum of concentrations of Ti$-$Ti pairs and
Ti$-$Ti$-V_\mathrm{Al}$ triples ($c_2$) at $T=T_m$ in the the conditions that correspond
to the point C, D, F, I, J of the Al$-$Ti$-$O$-$C phase diagram (Fig. \ref{f5}).}
\label{f19}
\end{figure}

Let us also discuss the possible scenario of violation of proportionality between the FOM
and $1/c_{\mathrm{Ti}^{4+}}$ even for samples annealed at the same temperature. We imply
that at rather low temperature the diffusion coefficient for certain defects is so small
that such defects can be considered as frozen ones. Let Ti ions be such defects while Al
vacancies remain mobile ones.
To calculate concentrations of Ti defects  we take into account two additional
constraints
\begin{eqnarray}
\label{301}
  c_{\mathrm{Ti}^{3+}}+ c_{\mathrm{Ti}^{4+}}+ c_{\mathrm{Ti}^{4+}-V_\mathrm{Al}^{3-}}&=&
   c_1=const, \\
 c_{\mathrm{Ti}^{3+}-\mathrm{Ti}^{3+}}+ c_{\mathrm{Ti}^{3+}-\mathrm{Ti}^{4+}}+
 c_{\mathrm{Ti}^{4+}-\mathrm{Ti}^{4+}-V_\mathrm{Al}^{3-}}+
 c_{\mathrm{Ti}^{3+}-\mathrm{Ti}^{4+}-V_\mathrm{Al}^{3-}}+
  c_{\mathrm{Ti}^{3+}-\mathrm{Ti}^{3+}-V_\mathrm{Al}^{3-}}&=& c_2=const. \label{302}
\end{eqnarray}
These constraints forbid formation of new Ti clusters  and dissociation of existing Ti
clusters. The concentrations of defects with one Ti ($c_1$) and with two Ti ions ($c_2$)
calculated for the same imaginary samples c, d, f, i and j are shown in Fig. \ref{f19}.
 For the system
with the constraints (\ref{301}) and (\ref{302}) the relation (\ref{4})  is not
satisfied. Therefore the relation (\ref{13}) is violated which can be demonstrated by
direct calculations.

We calculate equilibrium concentration of Ti$^{3+}$,  Ti$^{4+}$ and $\mathrm{Ti}^{3+}-
\mathrm{Ti}^{4+}$  under the constraints (\ref{301}) and (\ref{302}) at $T=1600$ K (we
imply that  the frozen regime is reached at this temperature).
 We fix
$\mu_\mathrm{O}$ that corresponds to the point C$_1$ (Table \ref{t5}). The quantities
$c_1$ and $c_2$ are fixed to ones as in the samples c, d, f, i and j. The results of the
calculations are shown in Fig. \ref{f18} by triangles  labeled as  c1, d1, f1, i1 and j1.
One can see that indeed $c_{\mathrm{Ti}^{3+}}/c_{\mathrm{Ti}^{3+}-\mathrm{Ti}^{4+}}$ is
not proportional to $1/c_{\mathrm{Ti}^{4+}}$. Comparing Fig. \ref{f18} and Fig. \ref{f19}
we find that $c_{\mathrm{Ti}^{3+}}/c_{\mathrm{Ti}^{3+}-\mathrm{Ti}^{4+}}$ correlates with
 the $c_1/c_2$ ratio. Note that the violation of the relation (\ref{13}) and more general relation (\ref{4})
 is caused solely by the additional constraints (\ref{301}) and (\ref{302}), and therefore
  this effect most probably
  holds true for a wide range of "freezing" temperatures.

\begin{figure}
\includegraphics[width=0.4\linewidth]{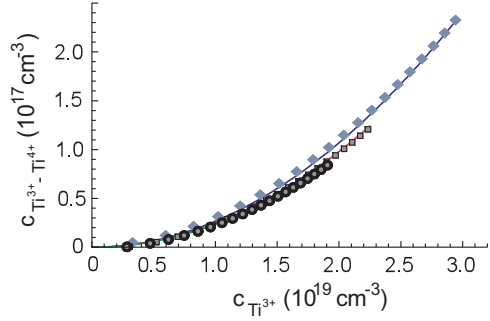}
\caption{Dependence of the concentration of $\mathrm{Ti}^{3+}-\mathrm{Ti}^{4+}$ on the
concentration of Ti$^{3+}$ calculated for the case where equilibrium concentrations are
reached at fixed total Ti concentration at $\mu_\mathrm{O}$=-10.0 eV and $T=T_m$
(diamonds), $T=2100$ K (squares) and $T=2000$ K (circles). Lines show the quadratic fit.
Total concentration of Ti is varied from $5\cdot 10^{18}$ cm$^{-3}$ to $10^{20}$
cm$^{-3}$} \label{f20}
\end{figure}

To conclude this analysis we discuss one more feature observed in \cite{m1}. It was found
in \cite{m1} that for some sets of samples the absorption coefficient $\alpha_{820}$ is
proportional to the square of the absorption coefficient $\alpha_{490}$. If NIR
absorption is caused by $\mathrm{Ti}^{3+}-\mathrm{Ti}^{4+}$ pairs such a behavior means
that the concentration of $\mathrm{Ti}^{3+}-\mathrm{Ti}^{4+}$ pairs is proportional to
the square of the concentration of $\mathrm{Ti}^{3+}$ isolated ions. It cannot be the
general property. Nevertheless our calculations show that for the samples annealed at the
same $\mu_\mathrm{O}$ this property is satisfied at least approximately. We specify
$\mu_{\mathrm{O}}=-10.0$ eV  and calculate equilibrium concentrations of defects   at
$T=T_m$,  $T=2100$ K and $T=2000$ K assuming that the total concentration of Ti
$c_{\mathrm{Ti}}$ is fixed. We consider 20 different $c_{\mathrm{Ti}}$ in the range from
$5\cdot 10^{18}$ cm$^{-3}$ to $10^{20}$ cm$^{-3}$. Obtained concentrations of
$\mathrm{Ti}^{3+}-\mathrm{Ti}^{4+}$ are plotted in Fig. \ref{f20} against the
concentration of $\mathrm{Ti}^{3+}$. One can see that for given $T$ the law
$c_{\mathrm{Ti}^{3+}-\mathrm{Ti}^{4+}}=\kappa c^2_{\mathrm{Ti}^{3+}}$ is satisfied. The
coefficient $\kappa$ weakly depends on temperature. Thus the observed in \cite{m1}
correspondence between $\alpha_{820}$ and $\alpha_{490}$ does not contradict the pair
model of NIR absorption.

\section{Conclusion}

In conclusion, with reference to Ti:doped sapphire we analyzed factors that determine the
charge state of dopants in insulating crystals.  Basing on DFT calculation combined with
thermochemical data we found that Ti atoms enter into the Al$_2$O$_3$ crystal
predominantly in the form of substitutional $\mathrm{Ti}^{3+}$ and $\mathrm{Ti}^{4+}$
ions. An essential fraction of such defects bind in pairs, triples etc., or form
complexes with Al vacancies.

 For laser application it is important to reduce the concentration of
$\mathrm{Ti}^{3+}-\mathrm{Ti}^{4+}$ pairs  keeping concentration of unpaired Ti$^{3+}$
ions at high level. To increase
$c_{\mathrm{Ti}^{3+}}/c_{\mathrm{Ti}^{3+}-\mathrm{Ti}^{4+}}$ ratio to the level of $10^3$
or higher the crystals should be grown or annealed in the reduced conditions.
Alternatively one can fix the total concentration of Ti at a relatively low level
($\lesssim 10^{18}$ cm$^{-3}$). Annealing in the conditions that correspond to
intermediate values of the oxygen chemical potential may result in a decrease of the
$c_{\mathrm{Ti}^{3+}}/c_{\mathrm{Ti}^{3+}-\mathrm{Ti}^{4+}}$ ratio.

We have shown that codopants that form charged defects may change relative concentrations
of the main dopants in different charge states. In particular, we have found that growth
or annealing of Ti:sapphire in the presence of nitrogen  compounds results in the
appearance of a large number of negatively charged nitrogen defects that cause  a
decrease of the $c_{\mathrm{Ti}^{3+}}/c_{\mathrm{Ti}^{3+}-\mathrm{Ti}^{4+}}$ ratio.

We have demonstrated that growth or annealing of doped crystals in the presence of
additional compounds may influence the charge state of dopants indirectly. Such a
situation is realized in Ti:sapphire grown or annealing in the presence of carbon or its
compounds. Carbon defects have large formation energies and their equilibrium
concentrations are quite low. Nevertheless the presence of carbon and its compounds may
increase the $c_{\mathrm{Ti}^{3+}}/c_{\mathrm{Ti}^{3+}-\mathrm{Ti}^{4+}}$ ratio due to
lowering of the oxygen chemical potential and increasing of the titanium chemical
potential.

We analyze the general relation between the concentrations of isolated and complex
defects. We find that normally the concentration of complex defects is proportional to
the product of concentrations of isolated defects which form the complex defect. The
coefficient of proportionality depends on the binding energy and on the temperature of
annealing or the melting temperature (for as-grown samples). This relation  is violated
if some defects are frozen. Such a situation is expected at rather low temperature of
annealing. In application to Ti:sapphire it means that the
$c_{\mathrm{Ti}^{3+}}/c_{\mathrm{Ti}^{3+}-\mathrm{Ti}^{4+}}$ ratio and FOM are
proportional for the inverse concentration of $\mathrm{Ti}^{4+}$ only for samples
annealed at the same temperature and even in that case such a proportionality can be
violated if the temperature of annealing is quite low.

\section*{Acknowledgment}

This work was performed using computational facilities of the Joint computational cluster
of the State Scientific Institution "Institute for Single Crystals" and the Institute for
Scintillation Materials of the National Academy of Sciences of Ukraine incorporated into
the Ukrainian National Grid.


\begin{thebibliography}{99}

\bibitem{my} L. Y. Kravchenko and D. V. Fil, Defect complexes in Ti-doped sapphire:
A first principles study, J. Appl. Phys. 123, 023104 (2018).

\bibitem{m}  P. F. Moulton, Ti-doped sapphire: tunable solid-state laser, Opt. News 8(6), 9 (1982).
\bibitem{1}  P. F. Moulton,
Spectroscopic and laser characteristics of Ti:Al2O3, J. Opt. Soc. Am. B 3(1), 125 (1986).
\bibitem{2} P. Lacovara, L. Esterowitz, and M. Kokta,
Growth, spectroscopy and lasing of titanium-doped sapphire, IEEE J. Quantum Electron.
21(10), 1614  (1985).
\bibitem{uv} W. C. Wong, D. S. McClure, S. A. Basun, M. R. Kokta. Charge-exchange processes
in titanium-doped sapphire crystals. I. Charge-exchange energies and titanium-bound
excitons, Phys. Rev. B. 51, 5682 (1995).
\bibitem{nizh} S. V. Nizhankovskii, N. S. Sidel'nikova, and V. V. Baranov,
 Optical absorption and color centers in large Ti: sapphire
crystals grown by horizontally directed crystallization under reducing conditions, Phys.
Solid State 57(4), 781 (2015).

\bibitem{2a} A. J . Strauss, R. E. Fahey, A. Sanchez, and R.
L. Aggarwal, Growth and characterization of Ti : A1203 crystals for laser applications,
Proc. SPIE 681, 62 (1987).
\bibitem{3a} A. Sanchez, A. J. Strauss, R. L. Aggarwal,  M. M. Stuppi, R. E. Fahey,
W. R. Rapoport, and C. P. Khattak, Crystal growth, spectroscopy, and laser
characteristics of Ti:Al$_2$O$_3$, IEEE J. Quantum Electron. 24, 995 (1988).
\bibitem{3} R. L. Aggarwal, A. Sanchez, M. M. Stuppi, R. E. Fahey, A. J. Strauss,
W. R. Rapoport, and C. P. Khattak, Residual infrared absorption in as-grown and annealed
crystals of Ti:Al2O3, IEEE J. Quantum Electron. 24, 1003 (1988).
\bibitem{4} W. R.
Rapoport and C. P. Khattak, Titanium sapphire laser characteristics, Appl. Opt. 27, 2677
 (1988).
\bibitem{pinto} J. F. Pinto, L. Esterowitz, G. H. Rosenblatt, M Kokta,
D, Peressini,  Improved Ti:sapphire laser performance with new high figure of merit
crystals. IEEE Journal of Quantum Electronics 30, 2612 (1994).
\bibitem{zha} F. X. Zha, J. H. Zhang, and S. D. Xia, Electronic structure of the ion
pair model for Ti:Al$_2$O$_3$, J. Phys.: Condens. Matter 6, 6497 (1994).

\bibitem{m1} P. F. Moulton, J. G. Cederberg, K. T.
Stevens, G. Foundos, M. Koselja, J. Preclikova, Characterization of absorption bands in
Ti:sapphire crystals, Optical Materials Express 9, 2216 (2019).

\bibitem{d1} K. Matsunaga, A. Nakamura, T. Yamamoto, and Y. Ikuhara,
First-principles study of defect energetics in titanium-doped alumina,
 Phys. Rev. B
68, 214102 (2003).
\bibitem{d2} K. Matsunaga, T. Mizoguchi, A. Nakamura, T. Yamamoto, and Y.
Ikuhara, Formation of titanium-solute clusters in alumina: A first-principles study,
Appl. Phys. Lett. 84, 4795 (2004).





\bibitem{z-n} S. B. Zhang, J. E. Northrup, Chemical potential dependence
of defect formation energies in GaAs: Application to Ga self-diffusion, Phys. Rev. Lett.
67, 2339 (1991).

\bibitem{frey}  C. Freysoldt, B. Grabowski, T. Hickel, J. Neugebauer,
 G. Kresse, A. Janotti, and C. G. Van de Walle,
 First-principles calculations for point defects in solids,
 Rev. Mod. Phys. 86, 253 (2014).

 \bibitem{si} J. M. Soler, E. Artacho, J. D. Gale, A. Garcia, J. Junquera, P. Ordejon, and
D. Sanchez-Portal, The SIESTA method for ab initio order-N materials simulation, J.
Phys.: Condens. Matter 14, 2745 (2002).

\bibitem{pp}  B. E. Tegner, G. J. Ackland, Pseudopotential errors in titanium,
Computational Materials Science, 52, 2 (2012)).

\bibitem{t-r1} K. Reuter and M. Scheffler, Composition, structure, and stability of
RuO$_2$ (110) as a function of oxygen pressure, Phys. Rev. B 65, 035406 (2001).

\bibitem{t-r2} K. Reuter and M. Scheffler, First-Principles Atomistic Thermodynamics
for Oxidation Catalysis: Surface Phase Diagrams and Catalytically Interesting Regions,
Phys. Rev. Lett. 90, 046103 (2003).


\bibitem{tct} M. W. Chase, Jr., NIST-JANAF Thermochemical Tables, 4th ed. (American
Chemical Society and American Institute of Physics, New York, 1998).

\bibitem{h09} N. D. M. Hine, K. Frensch, W. M. C. Foulkes, and M. W. Finnis,
Supercell size scaling of density functional theory formation energies of charged defects,
Phys.
Rev. B 79, 024112 (2009).

\bibitem{h10} N. D. M. Hine, P. D. Haynes, A. A. Mostofi, and M. C. Payne,
Linear-scaling density-functional simulations of charged point defects in Al$_2$O$_3$
using hierarchical sparse matrix algebra, The Journal of Chemical Physics 133, 114111
(2010).

\bibitem{mp} G. Makov and M. C. Payne, Periodic boundary conditions in ab initio
calculations, Phys. Rev. B 51, 4014 (1995).

\bibitem{komsa} H.-P. Komsa, T. T. Rantala, and A. Pasquarello, Finite-size supercell
correction schemes for charged defect calculations, Phys. Rev. B 86, 045112 (2012).

\bibitem{cn1} M. Choi, J. L. Lyons, A. Janotti,  C. G. Van de Walle, Impact of
carbon and nitrogen impurities in high-$\kappa$ dielectrics on metal-oxide-semiconductor
devices, Appl. Phys. Lett 102, 142902 (2013).

\bibitem{c2} J. Zhu, K. P. Muth, R. Pandey,
Stability and electronic properties of carbon in $\alpha$-Al$_2$O$_3$, Journal of Physics
and Chemistry of Solids 75, 379 (2014).

\bibitem{c3} L. Ao, Y.G. Yuan, Y. Tian, J.L. Nie, H.Y. Xiao, H. Chen, X. Xiang,
X.T. Zu, Electronic and magnetic properties of C-doped $\alpha -\mathrm{Al}_2\mathrm{O}3$
by DFT calculations, Computational Materials Science 110, 368 (2015).


\bibitem{nizh1} S. V. Nizhankovskiy, N. S. Sidelnikova, V. V. Baranov,
 Influence of crystal growth conditions and carbothermal treatment
 on activator charge state in Ti:sapphire,
 Functional Materials 25, 208 (2018).



\bibitem{lv1} J. Zhang, J. Ding, and Y. Zhang, Electronic and optical properties of
Ti$^{3+}$ doped $\alpha$-Al$_2$O$_3$ crystals: First-principles calculations, Solid State
Commun. 149, 1188 (2009).


\bibitem{lv2} M. G. Brik, Ab-initio studies of the electronic and optical
properties of Al2O3:Ti$^{3+}$ laser crystals, Phys. B 532, 178 (2018).
\end{thebibliography}
\end{document}